# Recycling of β-$Li_3PS_4$-based all-solid-state Li-ion batteries: Interactions of electrode materials and electrolyte in a dissolution-based separation process

Kerstin Wissel[a, b, *], Aaron Haben[c], Kathrin Küster[d], Ulrich Starke[d], Ralf Kautenburger[c], Wolfgang Ensinger[a] and Oliver Clemens[b]

[a] Technical University of Darmstadt, Institute for Materials Science, Materials Analysis, Alarich-Weiss-Straβe 2, 64287 Darmstadt, Germany

[b] University of Stuttgart, Institute for Materials Science, Chemical Materials Synthesis, Heisenbergstraβe 3, 70569 Stuttgart, Germany

[c] Saarland University, Inorganic Solid State Chemistry, Elemental Analysis, Campus C4 1, 66123 Saarbrücken, Germany

[d] Max Planck Institute for Solid State Research, Heisenbergstraße 1, 70569 Stuttgart, Germany

Corresponding Author:

Dr. Kerstin Wissel

Email: kerstin.wissel@imw.uni-stuttgart.de

Fax: +49 711 685 61963

## Abstract

All-solid-state batteries are currently developed at high pace and show a strong potential for market introduction within the next years. Though their performance has improved considerably over the last years, investigation of their sustainability and the development of suitable recycling strategies have received less attention. However, their potential for efficient circular processes must be accessed comprehensively. In this article, we investigate the separation of the solid electrolyte β-$Li_3PS_4$ from different lithium transition metal oxide electrode materials ($LiCoO_2$, $LiMn_2O_4$, $LiNi_{0.8}Mn_{0.1}Co_{0.1}O_2$, $LiFePO_4$, $LiNi_{0.85}Co_{0.1}Al_{0.05}O_2$ and $Li_4Ti_5O_{12}$) via an approach based on the dissolution and subsequent recrystallization of the thiophosphate using N-methylformamide as solvent. A combination of X-ray diffraction, scanning electron microscopy, energy-dispersive X-ray spectroscopy, inductively coupled plasma-mass spectrometry, iodometric titration and X-ray photoelectron spectroscopy as well as electrochemical impedance spectroscopy and electrochemical characterization was used to characterize the electrolyte and electrode materials before and after separation. We find that the presence of electrode materials in the dissolution process can lead to significant chemical reactions. These interactions can (but most not) lead to strong alteration of the electrochemical characteristics of the individual compounds. Thus, we show that an efficient recovery of materials will likely depend on the precise material combination within an all-solid-state battery.

# 1 Introduction

Since the first commercialization of rechargeable Li-ion batteries (LIBs), this technology has become omnipresent in our everyday lives. In recent years, the exponential growth of consumer electronics and stationary battery markets, coupled with the significant advancement in electronic vehicle technology, has led to a surge in the demand for LIBs - a trend expected to persist in the coming years. However, inherent safety concerns related to the use of combustible liquid electrolytes and limited energy density have also prompted the development of various next-generation battery technologies. Among these, all-solid-state batteries (ASSBs) utilizing solid electrolytes (SEs) are considered highly promising, and different material classes including oxides, polymers, halides and sulfides are intensively studied with respect to their potential use as SEs.[1-4] Owing to their high ionic conductivities in the range of $10^{-4}$ to $10^{-2}$ $S \cdot cm^{-1}$, sulfide-based electrolytes including thiophosphates such as Li-P-S (LPS)-based glasses or glass ceramics (e.g. $xLi_2S \bullet (100-x)P_2S_5$, $Li_7P_3S_{11}$, β-$Li_3PS_4$), argyrodites $Li_6PS_5X$ (X = Cl, Br, I), thio-LISICONs and $Li_{11-x}M_{2-x}P_{1+x}S_{12}$ (M = Ge, Sn, Si) have gained significant interest.[5]

With the increasing demand for batteries, the development of sustainable recovery and recycling strategies for spent batteries has become a matter of growing urgency. Commonly employed recycling methods for conventional LIBs encompass three types: pyrometallurgical recycling, hydrometallurgical recycling, and direct recycling.[6-9] Pyrometallurgical recycling entails subjecting the batteries to high-temperature treatments to break down the materials, while hydrometallurgical recycling involves dissolving the battery components to extract metals and other materials individually. Direct recycling, on the other hand, involves separating battery components without disassembling them into individual constituents, often aided by various organic solvent. The main focus of the recycling of LIBs is set on the recovery of valuable metals such as Li, Co and other transition metals.[10]

With the growing significance of ASSBs, it is crucial to recognize that recycling strategies employed for conventional LIBs may not be directly applicable to ASSBs due to their unique properties, primarily attributed to the chemical nature of the SEs employed in these batteries.[11] With respect to thiophosphate-based ASSBs, pyro- and/or hydrometallurgical recycling processes are not feasible since reactions with water or $O_2$ lead to a chemical degradation of the electrolyte. In general, the retention of the thiophosphate building units (e.g., ortho-thiophosphate units $PS_4^{3-}$, meta-thiodiphosphate units $P_2S_6^{2-}$, pyro-thiodiphosphate units $P_sS_7^{4-}$ etc.) during the recycling should be considered key. The economic value of thiophosphates lies not within the contained elements themselves but rather in the P-S bonds which can only be formed from elemental precursors using energy-intensive chemical synthesis approaches.[11] Related to the potential for solution-processing of thiophosphate electrolytes [12-20], direct recycling strategies involving the dissolution and subsequent recrystallization of the electrolyte might prove to be especially promising.[21] A first proof-of-principle of this dissolution-based separation strategy has been reported by Tan et al.[22] for a model system of Li|$Li_6PS_5Cl$|$LiCoO_2$ employing ethanol as solvent. They were able to separate the insoluble cathode material from the dissolved electrolyte. After the removal of the solvent, $Li_6PS_5Cl$ could be recovered, and the recycled materials could be used to build new batteries.

While $Li_6PS_5Cl$ can be dissolved in ethanol [23-25], investigation of the dissolution and recrystallization behavior of β-$Li_3PS_4$ have shown that ethanol leads to considerable decomposition of the SE, though both electrolytes contain $PS_4^{3-}$ units. For β-$Li_3PS_4$, we were able to show in a previous study[26] that the polar, weakly protic solvent N-methylformamide (NMF) is a suitable solvent allowing for a retention of $PS_4^{3-}$ units during the dissolution and recrystallization. The recrystallization proceeds via an intermediate phase $Li_3PS_4 \cdot 2NMF$. Thus, depending on the considered thiophosphate SE and the used solvent, complex reactions take place between the electrolyte and the solvent during the dissolution process. The possibilities for side reactions could increase due to the presence of different transition metal oxide (TMO) electrode materials in the dispersion formed after the dissolution of the electrolyte material. Commonly used electrode materials are for example layered materials such as $LiCoO_2$ (LCO), $LiNi_{0.8}Mn_{0.1}Co_{0.1}O_2$ (NMC811) and $LiNi_{0.85}Co_{0.1}Al_{0.05}O_2$, spinel-type $LiMn_2O_4$ (LMO) and $Li_4Ti_5O_{12}$ (LTO) and olivine-type $LiFePO_4$ (LFP).[27] Thiophosphates are known to have a low electrochemical stability window, which makes them unstable towards charged cathode and anode materials.[28] Already in the discharged state, these TMOs contain transition metal species with oxidation states higher than the preferred oxidation states of stable sulfides.[11] The presence of dissolved ions such as $PS_4^{3-}$ or $S^{2-}$ might in this context further enhance redox interactions under the formation of metal sulfides, polysulfides and/or sulfur leading to the degradation of the electrolyte and electrode material. Other known issues of TMOs[29] like irreversible layer-spinel-rock salt phase transformation and transition metal ion dissolution might also be influenced and could lower the feasibility of the recycling strategy.

The present study aims to investigate the nature of possible interactions between the dissolved β-$Li_3PS_4$ and different TMO electrode materials and their influence on the functional properties. In order to relate occurring reactions to a specific electrode material, systematic studies on mixtures of one of the electrode materials and β-$Li_3PS_4$ are conducted. After the dissolution in NMF, the separation of both fractions, the removal of the solvent and recrystallisation, structural changes of the recycled electrolyte and electrode materials are examined using X-ray diffraction (XRD) and Rietveld analysis. Detailed investigations of morphological and compositional modifications of the electrolyte and electrode fraction are conducted separately. For this, scanning electron microscopy (SEM), energy-dispersive X-ray spectroscopy (EDX), inductively coupled plasma-mass spectrometry (ICP-MS), iodometric titration and X-ray photoelectron spectroscopy (XPS) are used. Finally, the obtained results are related to the ionic and electronic conductivity of the recycled β-$Li_3PS_4$ and the electrochemical performance of the recycled electrode materials. This is used to derive information about compatibilities of different components within a solution-based recycling process, and can help to develop direct recycling strategies of thiophosphate-based ASSBs for an establishment of a circular economy.

# 2 Experimental

## 2.1 Material Preparation

The pristine electrolyte β-$Li_3PS_4$ was purchased from NEI Corporation (USA).

The pristine electrode materials $LiCoO_2$ (LCO), $LiMn_2O_4$ (LMO), $LiNi_{0.8}Mn_{0.1}Co_{0.1}O_2$ (NMC811) and $Li_4Ti_5O_{12}$ (LTO) were synthesized by solid-state reactions using $Li_2CO_3$ (AlfaAesar, 99%), $Co_3O_4$ (thermoscientific, 99.7%), $Mn_2O_3$ (Sigma-Aldrich, -325 mesh, 99%), NiO (Sigma-Aldrich, <50nm, 99.8%) and $TiO_2$ (thermoscientific, anatase, 99.6%, -325 mesh) as precursor materials. Mixtures of stoichiometric amounts of the respective oxides were prepared by intimately grinding using mortar and pestle and heated to 850 °C for 12 h in air, before regrinding and reheating at the same temperature for further 12 hours. The synthesis of $LiNi_{0.8}Co_{O.15}Al_{0.05}$ (NCA) was achieved using a method adopted from Qiu et al. [30]. For this, stoichiometric amounts of the metal acetate (hydrate) $Ni(CH_3COO)_2{\cdot}4H_2O$ (Sigma-Aldrich, 99.995 %), $Co(CH_3COO)_2$ (Sigma-Aldrich, 99.995 %) and $Al(CH_3COO)_2$ (Sigma-Aldrich, basic) were dissolved in deionized water. Excess oxalic acid $C_2H_2O_4$ (Fluka, anhydrous, ≥ 99%) was added and the mixture was stirred for 3 h, before drying at 80 °C. The obtained ternary metal oxalate powder was calcined at 450 °C for 6 h in air. $LiOH{\cdot}H_2O$ (Acros Organics, 98+%) was added in a molar ratio of Li:M = 1.05:1 by thoroughly mixing using mortar and pestle. To obtain the final product, this mixture was heated to 700 °C for 24 h under flowing oxygen. All synthesized electrode materials were directly transferred into an Ar-filled glovebox. $LiFePO_4$ (LFP) was purchased from Sigma-Aldrich (> 97 %, < 5 µm particle size).

Further material handling was carried out in Ar atmosphere.

## 2.2 Separation of β-$Li_3PS_4$ from electrode materials

For the dissolution of β-$Li_3PS_4$, N-methyl formamide (NMF, 99%, thermo scientific) was used as solvent. To remove water, it was dried over molecular sieve (3 Å, 20% m/v, Sigma-Aldrich). The molecular sieve was removed from the solvent after 72 h via filtration. To avoid any contamination from colloidal molecular sieve particles within the solvent, vacuum distillation was carried out in addition. A residual water content of 20 ppm was determined by Karl Fischer titration (Titrator Compact C10SX, Mettler-Toledo).

Pristine β-$Li_3PS_4$ and one of the pristine active electrode materials was mixed using mortar and pestle in a 1:1 weight ratio. Subsequently, NMF was added using a solid:liquid ratio of 50 mg:1ml. After 4 h of stirring, undissolved precipitates were separated from the solution using centrifugation (5000 RPM, 5 min, Hettich Zentrifugen EBA 21). The decanted solution was additionally filtered using a syringe filter (Whatman GD/X 25, pore size:0.45 µm). Undissolved precipitates were washed tree times using additional NMF.

The filtered solution and the washed precipitates were filled into Schlenk-flasks which were connected to a Schlenk line. The flask containing the solution was heated to 240 °C under vacuum (p ≈ 5-8 • $10^{-2}$ mbar) for 4 h to remove the excess solvent and to initiate the recrystallization of β-$Li_3PS_4$. For the precipitates, heating to 120 °C under vacuum was sufficient to remove remaining NMF.

As reference, one experiment was conducted in which only the electrolyte without addition of any electrode material was dissolved; solid:liquid ratio as well as the drying and recrystallisation procedure were unchanged. This sample is referred to as "recrystallized β-$Li_3PS_4$".

## 2.3 Characterization

### 2.3.1 X-ray Powder Diffraction and Rietveld analysis

XRD patterns were recorded on a Rigaku SmartLab in Bragg-Brentano geometry with Cu $K_\alpha$ radiation with a wavelength of 1.542 Å and a Hypix-3000 detector. Samples were measured inside low background air-tight sample holders (Rigaku), which were sealed inside an Ar-filled glovebox.

Analysis of diffraction data was performed via the Rietveld method with the program TOPAS V.6.0. For the crystal structure determination, a coupled Rietveld analysis of X-ray and neutron powder diffraction data was performed. The instrumental intensity distribution of the XRD and NPD instruments were determined empirically from a fundamental parameter set determined using a reference scan of $LaB_6$ (NIST 660a) and $Al_2O_3$ (NIST 676a), respectively. Microstructural parameters (i.e., crystallite size and strain broadening) were refined to adjust the peak shapes.

For the determination of amorphous phase contents, the samples were mixed in a defined weight ratio with $Al_2O_3$ (calcinated at 1100 °C) and XRD patterns were recorded. The calculation of the respective amorphous fraction was performed using the internal standard method as implemented in TOPAS V.6.0. The phase fraction of unidentified crystalline phase(s) was estimated based on degree of crystallinity determinations also implemented in the software.

### 2.3.2 Scanning electron microcopy and energy-dispersive X-ray spectroscopy

Scanning electron microscopy (SEM) images were recorded using a secondary electron detector of a Philips XL30-FEG microscope operating at 10 keV. Prior to the measurements, a layer of Au was sputtered onto the samples, which were placed on a carbon pad.

Due to a partial overlap of Au M and S and P K X-ray emission lines, energy-dispersive X-ray spectroscopy (EDX) measurements were conducted using an EDAX Genesis detector without an additional coating with gold. For each sample, three spots were measured. Quantitative analysis was performed using the EDAX Genesis software.

### 2.3.3 Inductively coupled plasma mass spectrometry

Quantitative elemental analysis of the electrolyte materials via inductively coupled plasma – mass spectrometry (ICP-MS) was conducted with an Agilent ICP-MS system 8900 with triple quadrupole ICP-QQQ and SPS4 autosampler. For the sample preparation, ~ 4mg of the respective electrode material was dissolved in 10 ml ultrapure water (0.055 µS/cm$^2$) (PURELAB® Chorus 1 ultrapure water filtration unit, Elga LabWater). A solution with 10 mg/l of Sc (1g/l in 5% HNO3, Alfa®) and Ho (1g/l in 2-3% $HNO_3$, Merck Certipur®) in ultrapure water was prepared as the internal standard stock solution for all ICP-MS measurements. $HNO_3$ (ROTIPURAN® Supra 69%, Carl Roth) was used to acidify the measurement solutions. Argon 5.0 (Ar ≥ 99.999 mol%, ALPHAGAZ™ 1 Argon, Air Liquide) was used as plasma gas for ICP-MS measurements. For calibration purpose Li (1 g/l in 0.5 M $HNO_3$, Merck Certipur®), Al (1 g/l in 0.5 M $HNO_3$, Bernd Kraft), P (1 g/l in water, AccuStandard AccuTrace®), Ti (1 g/l in water tr. HF, VWR), M (1 g/l in 0.5 M $HNO_3$, Fluka), Fe (1 g/l in 0.5 M $HNO_3$, Merck Certipur®), Co (1 g/l in 0.005 M $HNO_3$, Fluka) and Ni (1 g/l in 0.005 M $HNO_3$, Fluka) ICP-MS standards solutions were used. The measurements were performed with different reaction cell gas modes: $^{31}P$ was measured in $O_2$ reaction gas mode via mass pair ($Q_1$, $Q_2$) = (31, 47), $^{27}Al$, $^{47}Ti$, $^{55}Mn$, $^{56}Fe$, $^{59}Co$ and $^{60}Ni$ in He collision gas mode and $^{7}Li$ in “NoGas”-mode. An external calibration was done for quantification.

### 2.3.4 Electrochemical impedance spectroscopy

Electrochemical impedance spectroscopy (EIS) of the pristine as well as of the recycled electrolyte materials after separation from the electrode materials was performed using a TSC battery test cell (rhd instruments). Heating and cooling were achieved with a temperature-controlled cell stand. For this, 30 mg of each powder was uniaxially pressed into a free-standing pellet with a diameter of 7 mm (pelletizing pressure of 254 MPa for 5 min). The thickness of the obtained pellets was measured using a caliper. Conducting gold layers were sputter-coated onto the surfaces. Electrical impedance measurements were performed using an electrochemical impedance analyzer NEISYS (Novocontrol Technologies) in a frequency range between 1 MHz and 0.1 Hz with a root-mean-square amplitude of 10 mV in a temperature range between 20 and 80 °C (ΔT=5K, heating and cooling). Received data was analyzed using the software RelaxIS3 (rhd instruments GmbH & Co. KG).

### 2.3.5 Chronoamperometry

For chronoamperometry (CA) measurements pellets of the electrolyte materials with blocking gold electrodes were prepared as described above. A constant potential of 2 V was applied and the current-time response was recorded over a duration of 12 h to obtain a steady state current measurement. The measurements were performed with a VMP potentiostat (BioLogic Science Instruments).

### 2.3.6 Iodometric titration

For iodometric titration experiments, ~30 mg of a respective electrode material was dissolved in a mixture of 15 ml 20 % KI solution and 15 ml 1M HCl. The KI solution and HCl were freshly prepared using Ar-flushed water. Back titration was carried out using a 5mM $Na_2S_2O_3$ solution, 5 to 8 drops of 1 wt% starch solution were added as indicator. To minimize atmospheric oxidation, the experiments were performed under a flow of Ar.

### 2.3.7 X-ray photoelectron spectroscopy

For the preparation of XPS samples, double-sided adhesive tape, functioning in addition as an insulating layer, was used. Small quantity of powders of selected electrode materials were placed unto the tape which was previously stuck to the sample holder. Samples were transferred under Ar into the XPS chamber.

The X-ray photoelectron data were recorded using a Kratos AXIS Ultra spectrometer and monochromatized Al Kα radiation with an energy of 1486.6 eV. Survey spectra were acquired with a pass energy of 80 eV and detailed spectra with a pass energy of 20 eV.

The binding energies were calibrated to the C-C peak of adventitious carbon in the C 1s spectrum at 284.8 eV.

### 2.3.8 Electrochemical testing

For electrode preparation, 85 wt.% of pristine and recycled active electrode materials were mixed homogeneously with 5 wt.% Carbon Black Super P® (TIMCAL Ltd., Switzerland) as conducting additive and 10 wt.% polyvinylidene fluoride (Solef® PVDF, Solvay, Germany). The binder solution consisted of 10 wt.% PVDF in N-methyl-2-pyrolidone (NMP, BASF, Germany). The obtained slurries were tape-casted onto aluminum or copper foils. After tape casting, the printed electrodes were dried at 55 °C for 24 h. Round electrodes with a diameter of 7.8 mm were cut out and the weight of the electrodes was measured. The electrodes were dried under vacuum in a Büchi oven (Büchi glass oven B-585) at 80 °C for 24 h before being transferred without further contact with air into an argon-filled glove box for cell assembly.

For testing, cells of two-electrode Swagelok®-type set-up were assembled. Reference/counter electrodes with a diameter of 8mm were stamped out of Li foil (thickness 0.75 mm, diameter 5 mm, Sigma Aldrich). Glass fiber membranes (Whatman GF/C) were used as separator, which was soaked in 180 µl electrolyte. As electrolyte a solution of 1 M $LiPF_6$ in EC:DMC, ratio 1:1 (LP30, Merck KGaA, Germany) was used. The testing was performed with a VMP potentiostat (BioLogic Science Instruments). Cyclic voltammetry (CV) measurements were conducted in suitable potential ranges for the respective active electrode with a scan rate of 0.1 $mVs^{-1}$. Galvanostatic charging with potential limitation (GCPL) was performed using different C rates of C/10, C/5, C/2, C and 2C with charging/discharging currents calculated in relation to the active electrode mass excluding the mass of the carbon black and PVDF used in electrode preparation. The different constant currents were applied to evaluate the cycling stability and the rate performance.

# 3 Results and Discussions

After the separation of the dissolved $Li_3PS_4$ and the electrode materials, both fractions were dried under vacuum to remove the solvent NMF. To obtain a recrystallisation of the electrolyte to β-$Li_3PS_4$, heating parameter were chosen according to our previous findings on the recrystallisation behavior of β-$Li_3PS_4$.[26]

First indications for different reactivities between the electrode materials and the dissolved electrolyte can already be observed when comparing the colors of the solutions after the separation of the electrodes (Figure S1). The strongest color changes are found for the solution separated from LFP (dark greenish-brown color) and NCA (dark yellow). In contrast to this, the solution separated from LTO remains colorless. Such changes could be related to formation of polysulfides $Li_2S_x$ and, thus, decomposition of ortho-thiophosphate units $PS_4^{3-}$ might be indicated. Dissolved polysulfides are known to have a very complex influence on the color of solutions with a strong dependence on the value of x and the concentration of such species.[31-33] In addition, it should be noted that transition metal complexes can also affect the color of a solution significantly.[34] Even though, there are many factors (e.g., type of transition metal and ligand, as well as coordination numbers, geometries and symmetries of the transition metal complex) influencing the color, this points to a certain solubility of the transition metals from the electrode materials.

After the removal of the solvent, all recycled electrolyte and electrode materials are optically indistinguishable color compared to their pristine counterparts. The only exception is found for LFP for which a color change from grey for the pristine LFP to black for the recycled LFP is observed (Figure S2). This points already to compositional changes due to interactions with the dissolved $Li_3PS_4$ and/or dissolution of LFP.

## 3.1 X-ray diffraction study

To receive insights into possibly occurring interactions between the dissolved $Li_3PS_4$ and the electrode materials, an X-ray diffraction study was performed. Comparing the diffraction patterns of the recycled electrolytes and electrode materials received after drying (and recrystallisation) to the pristine materials, significant differences can be found (Figure 1). Quantitative phase analysis was conducted using the Rietveld method (see Table S1 and S2 for quantitative analysis and refined lattice parameters).

For the recrystallized and recycled electrolyte powders (Figure 1 a), the main phase observed after the separation procedure is β-$Li_3PS_4$. Compared to the pristine β-$Li_3PS_4$, a small contraction in the *a* lattice parameters (< 1%) are found. At the same time, the *b* lattice parameters increase in the same order of magnitude, resulting in overall comparable unit cell volumes between ~641 and 645 Å$^3$. Similar trends have been observed for nanoporous β-$Li_3PS_4$ (synthesized from the $Li_3PS_4$·3THF complex through annealing at 160 °C) upon heating to higher temperatures than needed for the synthesis and can be likely related to a gradual release of residual solvent or solvent decomposition products.[35]

Furthermore, additional reflections of unidentified phase(s) are found in all recycled electrolytes except for the electrolyte separated from LMO. Within the crystalline phase fraction, these phases account for up to ~12 wt.% with the highest weight fraction found for the electrolyte separated from NCA; LCO, NMC811, LFP and LTO contain between ~5 and 7 wt.% of these phases. Since the intensity ratios of these additional reflections differ from sample to sample, it is indicated the several impurity phases form. The formation of other thiophosphates with other thiophosphate units (e.g., $Li_2P_2S_6$, $Li_7P_3S_{11}$ or $Li_4P_2S_6$)[36-38] is not observed. Additionally, the impurity phase $Li_2S$ found in pristine β-$Li_3PS_4$ (~3 wt.%) is not found in the recycled materials indicating that it is e.g. consumed due to reactions with the electrode materials during the dissolution process and/or increases the Li and S contents in the amorphous matrix of the electrolyte fraction (possible also enabling the formation of the unknown crystalline phases(s) after recrystallization).

In general, a considerable increase of the amorphous phase fractions from ~8 % in the pristine β-$Li_3PS_4$, to ~35 % in the recrystallized β-$Li_3PS_4$ and to ~37 % to 56 % in the recycled β-$Li_3PS_4$ are observed suggesting a dependence of the recrystallisation behavior of recycled β-$Li_3PS_4$ on the electrode material the electrolyte was separated from (e.g., due to the partial dissolution of the electrode materials). For the recycled electrolytes, the lowest increase in amorphous phase fraction was observed for the electrolyte separated from LTO; the highest amorphization is found for LMO, LFP and NCA. Interestingly, for the electrolyte separated from LMO, a strong amorphization takes place, while no unidentified crystalline phases are formed. This might suggest different degradation reactions compared to the other electrolyte materials.

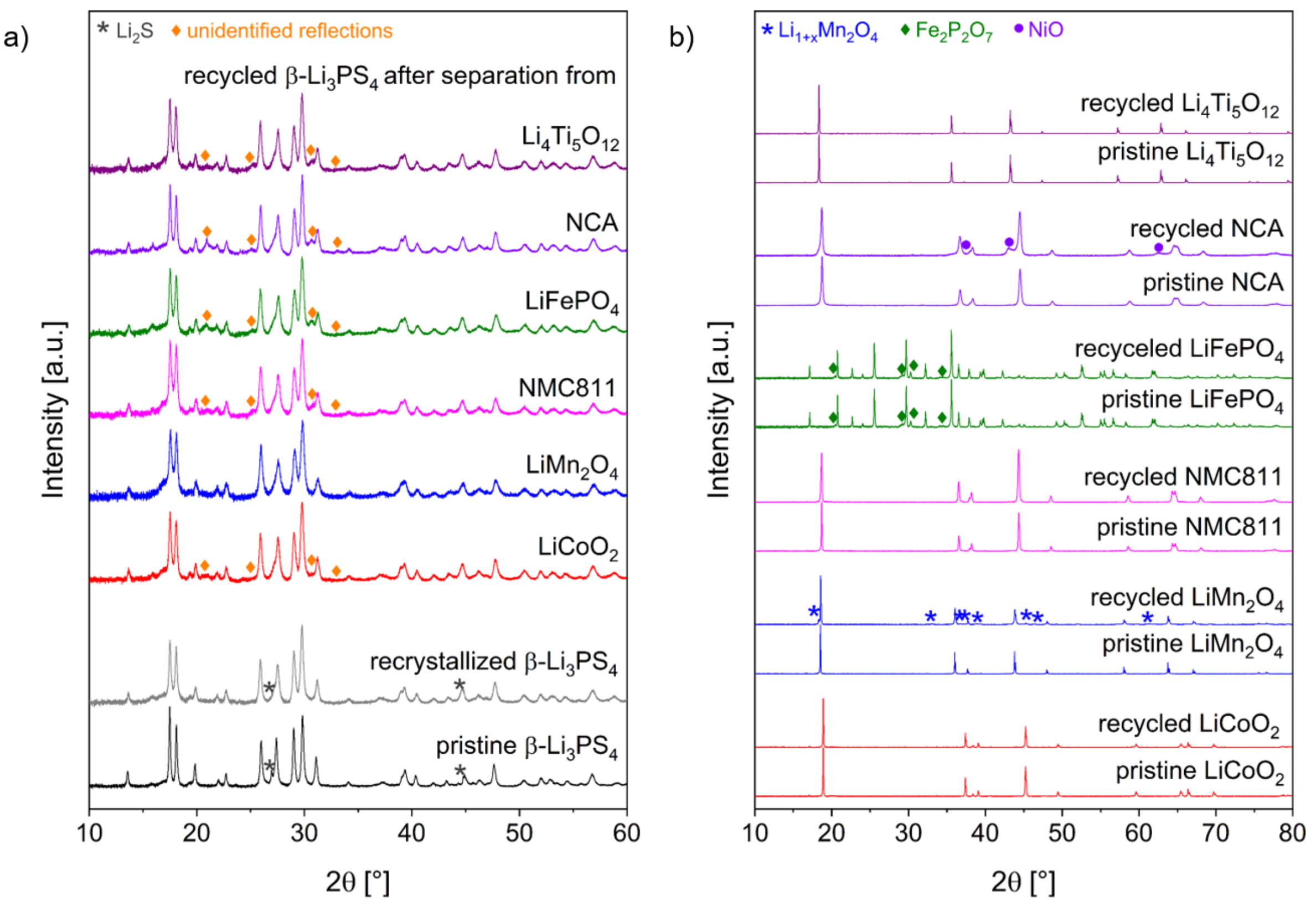

*Figure 1: XRD patterns of recycled β-$Li_3PS_4$ after separation from different electrode materials (a) and of recycled electrode materials (b) in comparison to the respective pristine materials.*

These observed changes can be partially related to the differences observed between the pristine and recycled electrode materials (Figure 1 b). While the XRD patterns of LCO, NMC811, LFP (phase fractions of impurity phase remain fairly constant in pristine and recycled LFP) and LTO are unchanged; additional crystalline phases can be found in the recycled LMO and NCA. Interestingly, the formation of crystalline transition metal sulfides is not observed in any of the XRD measurements. It should be noted that quantification errors in Rietveld analysis are in the order of 1–2 wt%. Thus, phase fractions of other crystalline phases might be too low to be detected (e.g., formation of a layer of metal sulfides on the surface of the electrode materials' particles).

In the case of the recycled LMO, ~17 wt.% of $Li_{1+x}Mn_2O_4$ ($x \leq 1$) are present besides $LiMn_2O_4$. This suggests that $LiMn_2O_4$ can be further lithiated due to the presence of $Li^+$ and $S^{2-}$ ions in solution (the latter acts as reductant). This leads, however, to a Li deficiency in the electrolyte fraction after separation. This implies that P- and S-rich phases might be formed, which could account to the high amorphous phase fraction of the recycled electrolyte. The absence of such Li-rich phases might also suggest that the unknown crystalline phase(s) found in the electrolytes separated from other electrodes might be Li-richer. The lithiation of the electrode material is correlated with an oxidation state change of the transition metal ion Mn from +III/IV in $LiMn_2O_4$ to +III in $Li_2Mn_2O_4$. The spinel-type $LiMn_2O_4$ represents a semi-discharged state of lithium manganese oxide cathode materials, while $Li_2Mn_2O_4$ can be considered fully discharged. Thus, such interactions can result in significant changes and degradation of both the electrolyte and the electrode material.

For recycled NCA, another degradation mechanism under the formation of ~17 wt.% NiO-like rock-salt phase is observed. Such instabilities can be related to the reduction tendency of $Ni^{3+}$ to $Ni^{2+}$ in Ni-rich cathodes (compare also degradation in ambient air)[39, 40]. The formation of NiO is only possible considering significant structural transformations of the electrode material connected to the inherent structural instability of layered oxides. This has been also observed in the context of cation-mixing which results in a gradual transformation of the layered structure into a spinel-like structure and finally into the rock-salt phase. This takes typically place starting on the surface and is gradually permeates into the bulk. [41, 42] The presence of $PS_4^{3-}$ and/or $S^{2-}$ in the solution might accelerate this degradation process. In particular, sulfide ions might lead to the stabilization of the preferred stable +II oxidation state of Ni under the formation of NiS. Reactions between NCA and the dissolved electrolyte would also imply that the recycled electrolyte would undergo a certain degradation due to a partial transformation of Li-P-S bonds into Li-P-O bonds. Such phases might contribute to the high amorphous phase fraction found in the recycled electrolyte separated from NCA. A similar behavior could be expected for NMC811 since it is also Ni-rich. Interestingly, the formation of crystalline NiO was not observed for the recycled NMC811. This could be correlated to the different particle sizes and surface areas of NMC811 and NCA (see also Figure 4 in section 3.3).

## 3.2 Detailed investigation of electrolytes

As was described in section 3.1, interactions between the dissolved electrolyte and the electrode materials do not only result in the formation of crystalline phases but also lead to significant amorphization. Thus, the identification of changes based on X-ray diffraction alone can be difficult. Special attention has to be given to the amorphous phase fractions present in the samples which increase considerably compared to pristine β-$Li_3PS_4$. Due consideration has to be given to other occurring morphological and compositional changes since these can significantly impact the functional properties of the recycled β-$Li_3PS_4$. Therefore, additional studies on such changes are required to find correlations to the ionic and electronic conductivities of the recycled electrolytes. With respect to ionic conductivity, it has been shown that higher crystallinity in $Li_3PS_4$ does not necessarily lead to improved properties [43, 44], however, even subtle changes in the chemical composition can affect structural properties and with this the transport properties significantly. Investigating the detailed composition might allow drawing conclusions on influencing factors.

To investigate morphology changes between the pristine and recycled β-$Li_3PS_4$ after separation from different electrode materials, SE micrographs were recorded (Figure 2). In agreement with our previous study on the recrystallisation of β-$Li_3PS_4$ from NMF [26], the morphologies of pristine and recycled β-$Li_3PS_4$ differ significantly. Compared to pristine β-$Li_3PS_4$, the recycled β-$Li_3PS_4$ powders consist of irregularly-shaped particles with a wider particle size distribution and overall smaller particles. The larger particles in the recycled electrolytes appear to be flake-like and splintery. Closest resemblance might be found between the pristine β-$Li_3PS_4$ and recycled β-$Li_3PS_4$ separated from LTO. Such changed particle shapes and, in particular, smaller particle sizes could considerably influence the ionic conductivity of the recycled β-$Li_3PS_4$.

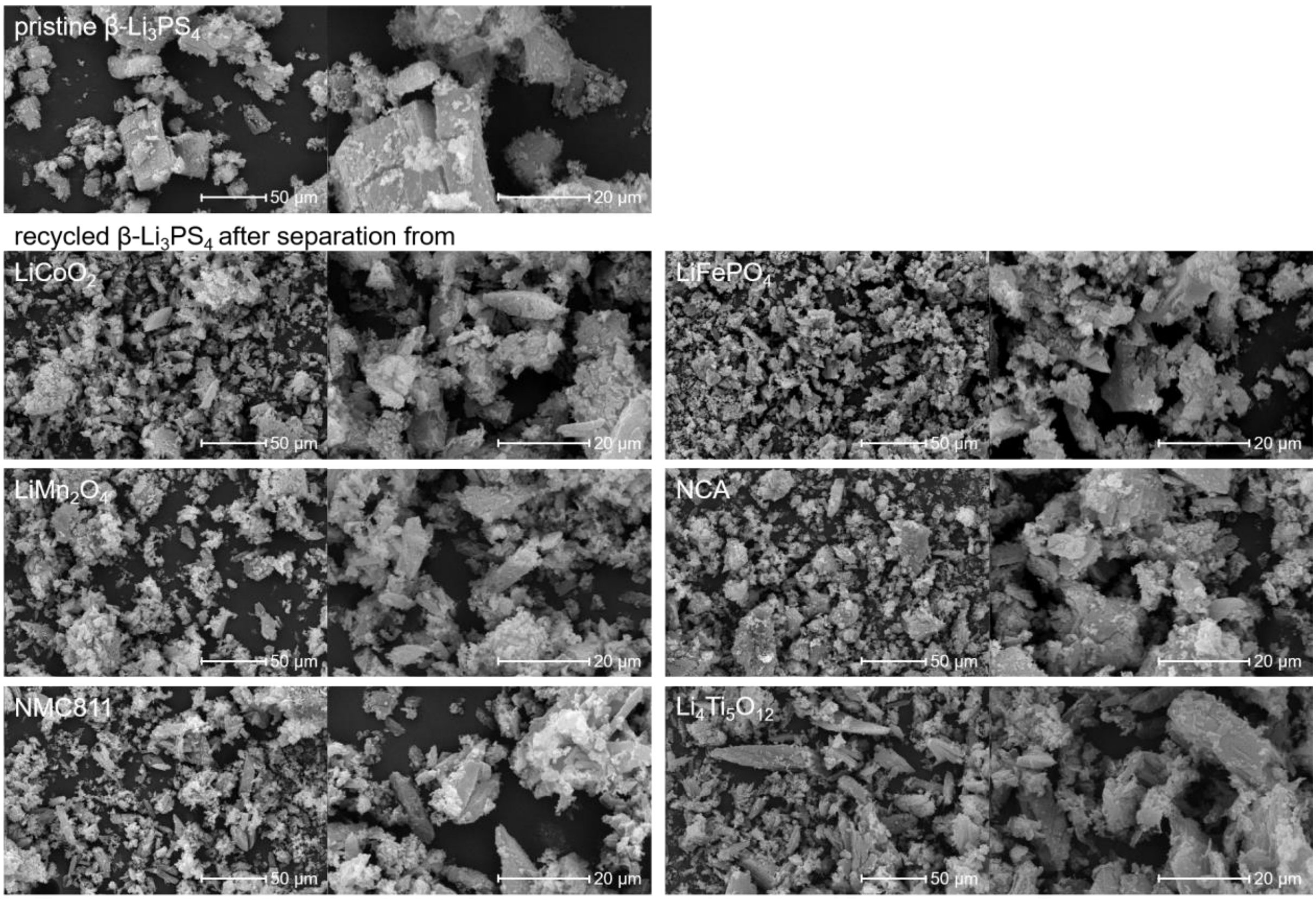

*Figure 2: SE micrographs of pristine and recycled β-$Li_3PS_4$ after separation from different electrode materials.*

Composition changes within the recycled electrolytes with a focus on possible transfer of transition metals due to undesired dissolution of the electrode materials was examined using ICP-MS measurements (EDX could not give meaningful insights due to the relatively high limit of detection of this method not suitable for trace analytes). A partial dissolution of transition metal ions could not only have significant impact on the electrode material itself, but also on the electrolyte fraction. Here, a dependence on the concentration of the transition metal, the type of phase formed incorporating the transition metal and its properties, as well as the distribution of this phase within the electrolyte phase can be expected. Furthermore, with respect to amorphous phase fractions, which can significantly influence the transport properties of thiosulfate electrolytes[43, 44], two aspects have to be considered: (i) The recrystallization behavior of β-$Li_3PS_4$ after its dissolution could be considerably changed due the presence of transition metal ions in the solution; (ii) the formation of the phase containing the transition metal might possibly result in elemental deficiencies within the electrolyte fraction with formal composition of Li:P:S of 3:1:4, hindering the formation of crystalline β-$Li_3PS_4$ and affecting the short-range order.

The molar ratios of Li, Al, Ti, Mn, Fe, Co and Ni to P based on ICP-MS measurements are summarized in Table 1 (molar concentrations are provided in Table S3). Owing to problems with the precise quantification of sulfur, no sulfur concentrations are reported. The Li:P molar ratio is in all materials close to 3 and comparable values between the pristine and recycled electrolytes are found. Slight deviations might relate to Li (or P) transfer between the electrode and electrolyte materials. Compared to the pristine β-$Li_3PS_4$, increased concentrations of Mn, Co and Ni can be observed in the recycled electrolytes though overall comparatively low elemental transfer are found. The highest increase in concentration is found for Mn in the electrolyte separated from LMO. With respect to this dissolution, the formation of NMF-soluble MnS under the stabilization of $Mn^{2+}$ might play a role. For the electrolytes separated from NMC811 and NCA increases in the Ni concentrations are observed; again, stronger reactivities are found for the electrolyte separated from NCA. The concentrations of the Mn and Co are similar to what is found in pristine β-$Li_3PS_4$. It should be noted that the concentrations of Al and Fe are below the limits of quantifications (marked with ND ("not determined")); the only exception is found for β-$Li_3PS_4$ separated from LFP showing increased Fe concentrations. Low Ti concentrations in the electrolyte separated from LTO point again to the high stability of LTO during the dissolution and separation process.

*Table 1: Molar ratios of Li, Al, Ti, Mn, Fe, Co and Ni to P of pristine and recycled β-$Li_3PS_4$ after separation from different electrode materials based on quantitative ICP-MS analyses. S contents were not determined.*

| | **Molar ratio** | | | | | | |
|---|---|---|---|---|---|---|---|
| | **Li:P** | **Al:P** | **Ti:P** | **Mn:P** | **Fe:P** | **Co:P** | **Ni:P** |
| **pristine β-$Li_3PS_4$** | 3.09 | ND | $5.15 \cdot 10^{-5}$ | $1.76 \cdot 10^{-5}$ | ND | $1.01 \cdot 10^{-6}$ | $2.74 \cdot 10^{-5}$ |
| **β-$Li_3PS_4$ separated from** | | | | | | | |
| **$LiCoO_2$** | 3.14 | ND | $2.89 \cdot 10^{-5}$ | $7.66 \cdot 10^{-6}$ | ND | $3.00 \cdot 10^{-6}$ | ND |
| **$LiMn_2O_4$** | 3.03 | ND | $3.24 \cdot 10^{-5}$ | $1.35 \cdot 10^{-3}$ | ND | $1.23 \cdot 10^{-7}$ | $8.50 \cdot 10^{-6}$ |
| **NMC811** | 3.22 | ND | $1.98 \cdot 10^{-5}$ | $3.92 \cdot 10^{-5}$ | ND | $4.98 \cdot 10^{-7}$ | $6.89 \cdot 10^{-5}$ |
| **$LiFePO_4$** | 3.03 | ND | $2.68 \cdot 10^{-5}$ | $1.38 \cdot 10^{-5}$ | $6.61 \cdot 10^{-4}$ | $3.41 \cdot 10^{-7}$ | $1.58 \cdot 10^{-5}$ |
| **NCA** | 3.09 | ND | $1.62 \cdot 10^{-5}$ | $2.67 \cdot 10^{-5}$ | ND | $7.45 \cdot 10^{-6}$ | $5.14 \cdot 10^{-4}$ |
| **$Li_4Ti_5O_{12}$** | 3.10 | ND | $3.54 \cdot 10^{-5}$ | $1.18 \cdot 10^{-5}$ | ND | $9.02 \cdot 10^{-8}$ | ND |

As already noted these morphological and compositional changes can be expected to have an influence on the functional properties (e.g., ionic and electronic conductivity) of the recycled electrolyte materials. Therefore, electrochemical impedance spectroscopy and chronoamperometry measurements were performed. The Nyquist plots of pristine, recrystallized and recycled β-$Li_3PS_4$ separated from the different electrode materials (measured at 25 °C) is shown in Figure 3 a). It was found that one parallel R/CPE element (R=resistor, CPE=constant phase element) connected in series to a CPE can be used for fitting the Li-ion transport process and the electrode-ion-blocking effect at the electrode; for recycled β-$Li_3PS_4$ separated from NCA, a model with two parallel R/CPE elements in series to a CPE is required. From the fits, the RT conductivity values and activation energies were determined (Figure 3 b); the Arrhenius plots are provided in Figure S3 a). The Nyquist and Bode (compare also Figure S4) plots of all samples are fairly similar with the exception of β-$Li_3PS_4$ separated from NCA and LFP. This shows that the formation of impurity phases, the increase of amorphous phase fractions as well as the partial dissolution of transition metal ions do not necessarily lead to a significant change of the impedance response (depending on the overall amounts of these phases/). The RT conductivities are in the same order of magnitude after the separation. The activation energies of the recrystallized and recycled electrolyte materials are comparable to those of other β-$Li_3PS_4$ synthesized via solvent-based preparation approaches.[31, 45, 46 47]

In contrast, the sample separated from LFP shows a considerable increase of the resistance, while the activation energy decreases significantly suggesting considerable modification of the Li-ion transport properties due to the observed interactions between the electrolyte and electrode material. For β-$Li_3PS_4$ separated from NCA, a significantly different frequency dependence of the impedance and phase angle (compared also Figure S3b and S4) and a strong decrease of the conductivity is observed. For both samples, the increasing resistances can be in principle correlated to the powder diffraction analysis which reveals not only the increased formation of amorphous phases compared to other electrode materials, but also the presence of a comparatively large phase fraction of presumably more resistive, unidentified phase(s). The strong interreactivity of dissolved $Li_3PS_4$ and these samples is also indicated by strong color change of the $Li_3PS_4$ containing NMF solution obtained after separation from LFP as well as of recycled LFP and the strong degradation of NCA under the formation of NiO, respectively. In addition, different elemental impurities are indicated in both samples. Further, LFP and NCA show a very different redox behavior after separation (see section 3.3).

Remarkably, it should be emphasized that the observed distinct redox reaction between LMO and β-$Li_3PS_4$ does not result in a significant impact on the conductivity. This indicates that side reactions are complex. The detailed nature of additional phases formed depends strongly on the electrode material and the interactions can affect the conductivity to different extents.

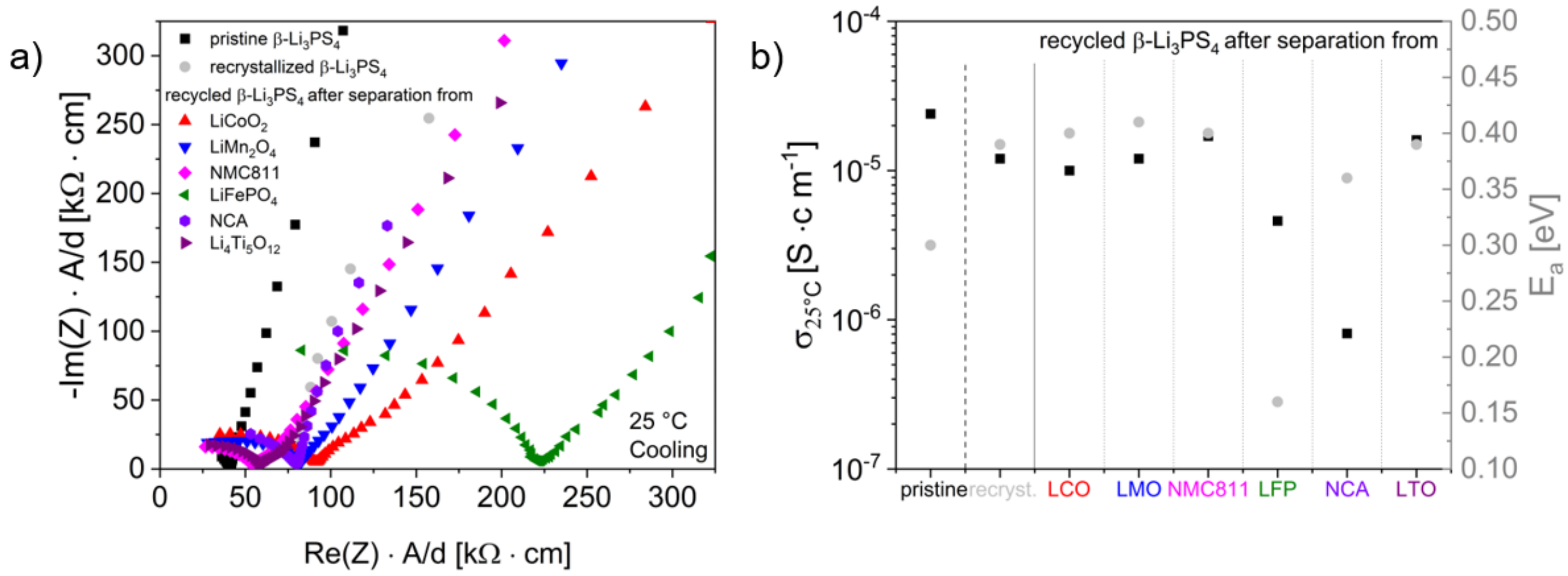

*Figure 3: a) Nyquist plots of pristine, recrystallized and recycled β-$Li_3PS_4$ after separation from different electrode materials measured at 25 °C. The extended spectrum of recycled β-$Li_3PS_4$ after separation from NCA is given in Figure S3; b) ionic conductivities $\sigma_{25°C}$ and activation energies $E_a$ of pristine, recrystallized and recycled β-$Li_3PS_4$ after separation from different electrode materials.*

The electronic conductivity of the recycled β-$Li_3PS_4$ was measured to be between ~ 7.8 x $10^{-10}$ and 1.3 x $10^{-9}$ S $cm^{-1}$ by chronoamperometry (Figure S5). The pristine and recrystallized β-$Li_3PS_4$ have comparable electronic conductivities of ~ 9.7 x $10^{-10}$ and 4.9 x $10^{-10}$ S $cm^{-1}$, respectively. These values are consistent with previously reported electronic conductivities for thiophosphate solid electrolytes [48-50], and the electronic contribution to the total conductivity is negligibly low [51]. Thus, the separation-process does not lead to an increased electronic conductivity.

## 3.3 Detailed investigation of electrode materials

Similar to the recycled electrolytes, morphological and composition changes can have significant influence on electrochemical properties of the recycled electrode materials. Due to the comparably low solubility of the investigated electrode materials, reactions are very likely to predominantly take place between the electrolyte ions in solution and the particle surfaces of the electrode materials, potentially leading to the formation of interfacial reaction products. Reactions occurring on the electrode materials' particle surfaces play a significant role since they can lead to the formation of a protective but also passivating layers on the particles. For example, modifications of electrode materials using sulfur, polysulfides or $Li_2S$ additives have been shown to improve the electrochemical performance due to the formation of artificial cathode electrolyte interphases.[52-54] Even though initial findings on compositional changes could be made based on X-ray diffraction, detailed investigations on e.g. oxidation state and overall composition changes are required. In particular, surface-sensitive characterization techniques like XPS can give important information with regards to changes of surface compositions.

The morphology of the pristine and recycled electrode materials was examined using SE microscopy (Figure 4). The shape and the size of the particles of the pristine electrode materials correspond to accordingly synthesized materials reported in literature [30, 55-59] Compared to the materials prepared via solid-state reactions, significantly smaller particle sizes are found for NCA synthesized using a wet-chemical approach. All materials show particle agglomeration. No difference between the pristine and recycled materials is observed indicating that the electrode materials remain relatively unchanged in terms of morphology.

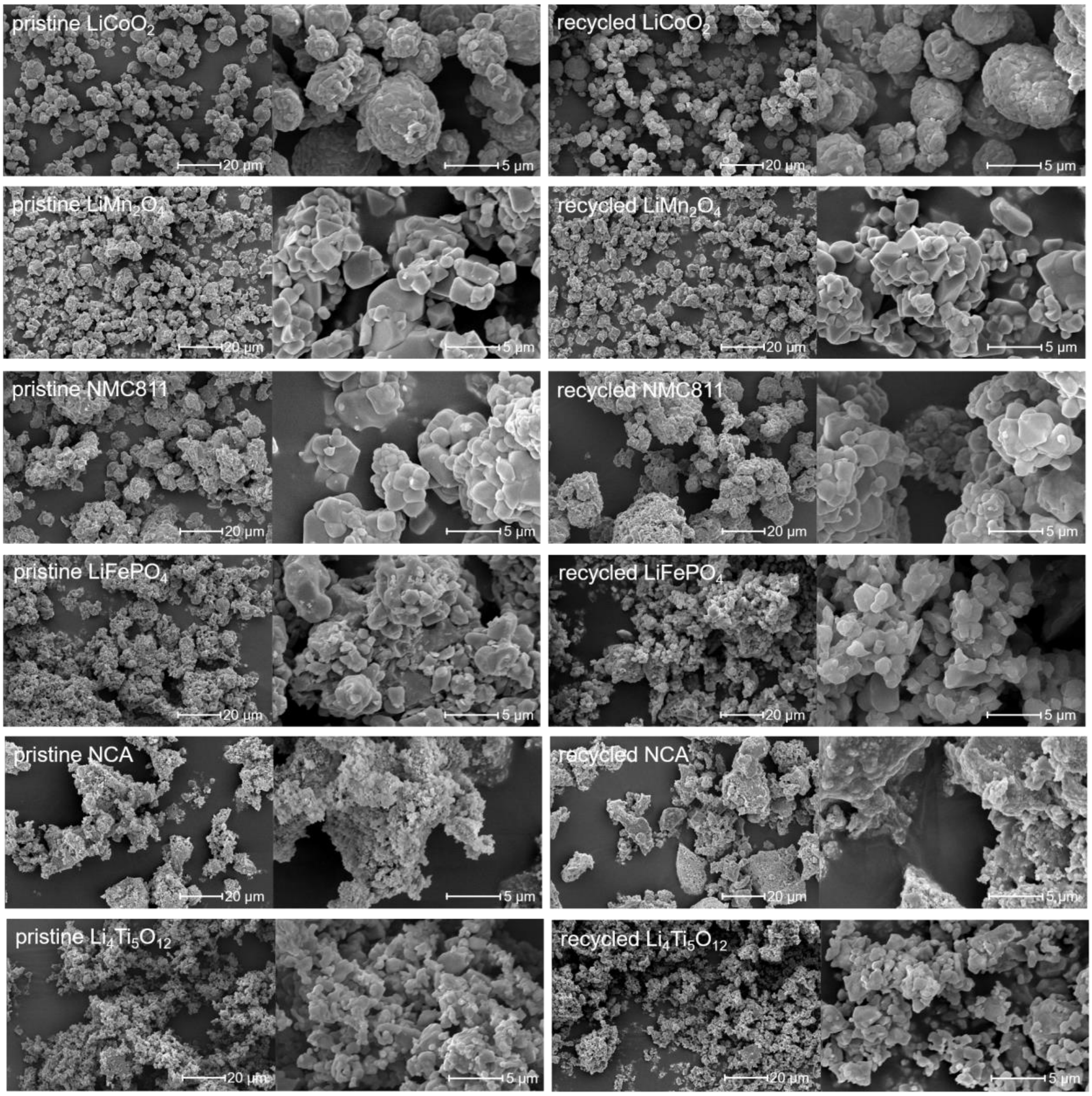

*Figure 4: SE micrographs of pristine and recycled electrode materials.*

For the determination of the chemical composition of the pristine and recycled electrode materials, EDX measurements were performed (Table S4). Within the error of this technique, the compositions of the electrode materials could be confirmed (in agreement with additional ICP-MS measurements). For the recycled electrode materials, additional signals corresponding to P and S are found, pointing towards carryover of these elements (presumably primarily on the particle surfaces) into the electrode material fraction after the separation. For recycled LCO, LMO, NMC811 (one deviating measurement might indicate compositional inhomogeneities) and LTO, these signals account for < 1 at.% of P and S. Recycled NCA shows higher amounts of P and S with up to ~ 2 at.% of P and ~ 4 at.% of S. For recycled LFP, overall lower P contents compared to the pristine materials might suggest a partial transfer of P (or $PO_4^{3-}$ units) into the solution and with this into the electrolyte fraction; S amounts are comparable to the values observed for recycled LCO, LMO, NMC811 and LTO.

X-ray diffraction of the recycled electrode materials (section 3.1) reveals the formation of crystalline $Li_{1+x}Mn_2O_4$ and NiO in the case of recycled LMO and NCA, respectively. Relatable changes of the average oxidation states of the transition metal cations of the electrode materials were determined using iodometric titration (Table 2). For LMO and NCA, lowered oxidation states compared to the pristine materials are observed, which correspond well to the expected degree of lowering with respect to the quantitative Rietveld analysis and the determined compositions. All other electrode materials (the oxidation state of LTO could not be determined due to its low solubility; its white color indicates, however, the presence of $Ti^{4+}$ [60]) possess average oxidation states close to the ideal oxidation states expected based on their composition.

*Table 2: Comparison of ideal and measured average oxidation states of the transition metal(s) (TM(s)) of pristine and recycled electrode materials determined by iodometric titration. *: For $LiFePO_4$, no significant color change of the solution after addition of KI and/or starch solution was observed, indicating an Fe oxidation state very close to +2.*

| | ideal average oxidation state of TM(s) | determined average oxidation state of TM(s) |
|---|---|---|
| **pristine $LiCoO_2$** | 3 | 2.89 |
| **recycled $LiCoO_2$** | | 2.87 |
| **pristine $LiMn_2O_4$** | 3.5 | 3.43 |
| **recycled $LiMn_2O_4$** | | 3.36 |
| **pristine NMC811** | 3 | 2.81 |
| **recycled NMC811** | | 2.78 |
| **pristine $LiFePO_4$** | 2 | 2.05* |
| **recycled $LiFePO_4$** | | 2* |
| **pristine NCA** | 3 | 3.02 |
| **recycled NCA** | | 2.73 |

To qualitatively investigate differences of surface compositions of the electrode materials before and after the dissolution-based separation process, XPS measurements were performed. As shown above, the strongest interactions between dissolved β-$Li_3PS_4$ and the electrode materials were found for the electrolyte fractions separated from LFP and NCA and/or the electrode materials LMO, LFP and NCA. Therefore, in the following, a focus is set on the investigation of these electrode materials. The 2p XP spectra of the respective main transition metal (Mn 2p for LMO, Fe 2p for LFP and Ni 2p for NCA) as well as the P 2p and of S 2p spectra are given in Figure 5. Additional spectra (e.g. O 1s and C1s) are provided in Figure S6. Interestingly, though the measurement parameters were identical, a comparison between the spectra of the main transition metals of the pristine and recycled electrode materials reveals significant differences in terms of counts per seconds (please consider the different scaling of the y axis of the pristine and recycled spectra) which is especially pronounced for LFP and NCA. This might indicate that the surfaces of the recycled LFP and NCA particles have a Fe and Ni deficiency, respectively. In contrast to this, the intensities of the Mn 2p spectra of pristine and recycled LMO are comparable.

For LMO, the comparison between the Mn 2p spectra of the pristine and recycled electrode material (Figure 5 a) reveals a significant shift of the Mn $2p_{3/2}$ and $2p_{1/2}$ signals towards lower binding energies, indicated by the shift of the center of gravity of the signals. The exact quantification of the oxidation state of Mn based on the Mn 2p spectrum is complicated due to the close proximity of $Mn^{4+}$, $Mn^{3+}$ and $Mn^{2+}$ emission lines and their multiple splitting. [61-63] However, the position of the Mn $2p_{3/2}$ between ~ 640 and 644 eV indicates a mixture of Mn in +2, +3 and +4 oxidation states. This result is in good agreement with our previous findings since the observed lithiation of $LiMn_2O_4$ to $Li_{1+x}Mn_2O_4$ leads to a lowering of the overall oxidation state towards +3. In addition, the increase of the intensity of the shoulder at ~ 640.9 eV in the recycled LMO suggests an increased contribution of $Mn^{2+}$.[64] This lower surface oxidation state might be related to the formation of MnS on the particle surfaces, reported at binding energies of 640.9 eV.[65]

The P 2p spectra of the pristine and recycled LMO differ significantly. While no P-containing species are found on the pristine material, a P 2p signal is found in the recycled LMO. The binding energy of the main signal of P–S bonds in the $PS_4^{3-}$ units of β-$Li_3PS_4$, which could be expected to be present on the particles after the separation process, is reported to be at 132.0 eV. [66-68] However, the center of gravity of the P 2p signal is shifted towards higher binding energies suggesting a considerable formation of e.g. P–$[S]_n$–P as well as oxygenated phosphorous species (phosphates, metaphosphates or $PS_{4-x}O_x^{3-}$)[69, 70].

The formation of sulfidic species on the particle surfaces of recycled LMO is indicated based on the S 2p spectrum. Fitting of the spectrum (fit not shown, see also shoulder at ~ 164.3 eV) suggests that at least two different doublet contributions at energies of ~ 161.7 and 162.9 eV and of ~ 163.1 and 164.3 eV, respectively, are required. An unambiguous assignment of the signals found at lower binding energies is not possible since different sulfidic species such as e.g., MnS (reported $2p_{3/2}$ binding energies between 161.4 and 162 eV)[71, 72] and $MnS_2$ (reported $2p_{3/2}$ binding energy ~161.9)[73], but also $Li_3PS_4$[66, 74] could be present impeding the deconvolution; the doublet signal at higher binding energies can be assigned to P–$[S]_n$–P species.[75] It should be noted that the spectrum of pristine LMO features a signal at binding energies characteristic for metal sulfates.[76] This might be related to reactions between the freshly synthesized LMO and gaseous sulfur species after the transfer into the glovebox in which numerous sulfur-containing compounds are stored. After the separation process, these metal sulfates are not present on the recycled materials particles.

For LFP and NCA in general similar observations can be made. The comparison between the Fe 2p spectra of the pristine and recycled LFP shows, however, in agreement with the previously reported results no significant change of the binding energy of the Fe $2p_{3/2}$ and $2p_{1/2}$ signals and, thus, the oxidation state of Fe remains close to +2 after the separation. Instead an additional shoulder is found in the spectrum of the recycled LFP at binding energies of ~707.0 eV, which corresponds well to pyrite $FeS_2$.[73] The presence of pyrite within the recycled LFP might also explain the observed color change after the separation.[77] The P 2p signal of both samples are very similar and confirm the predominant presence of phosphate species $PO_4^{3-}$ of $LiFePO_4$.[78, 79] The S 2p spectrum for recycled LFP confirms again the formation of sulfidic species on the particles' surfaces. [66, 73, 74, 80]

For NCA, in accordance with the previous results, the shift towards lower binding energies confirms a decrease of the Ni oxidation state related to the decomposition of the layered oxide and the formation of NiO.[81, 82] Compared to the P 2p spectra of recycled LMO and LFP, the spectrum of recycled NCA is slightly shifted towards lower binding energies which implies the presence of a higher fraction of P–S bonds of β-$Li_3PS_4$ instead of P–$[S]_n$–P and oxygenated phosphorous species. In addition, the S 2p signal has a smaller width indicating lesser P–$[S]_n$–P species. The XPS measurements of NMC811 (Figure S7) shows similar trends compared to NCA. However, in agreement with the previous findings, the decomposition under formation of NiO is significantly less pronounced.

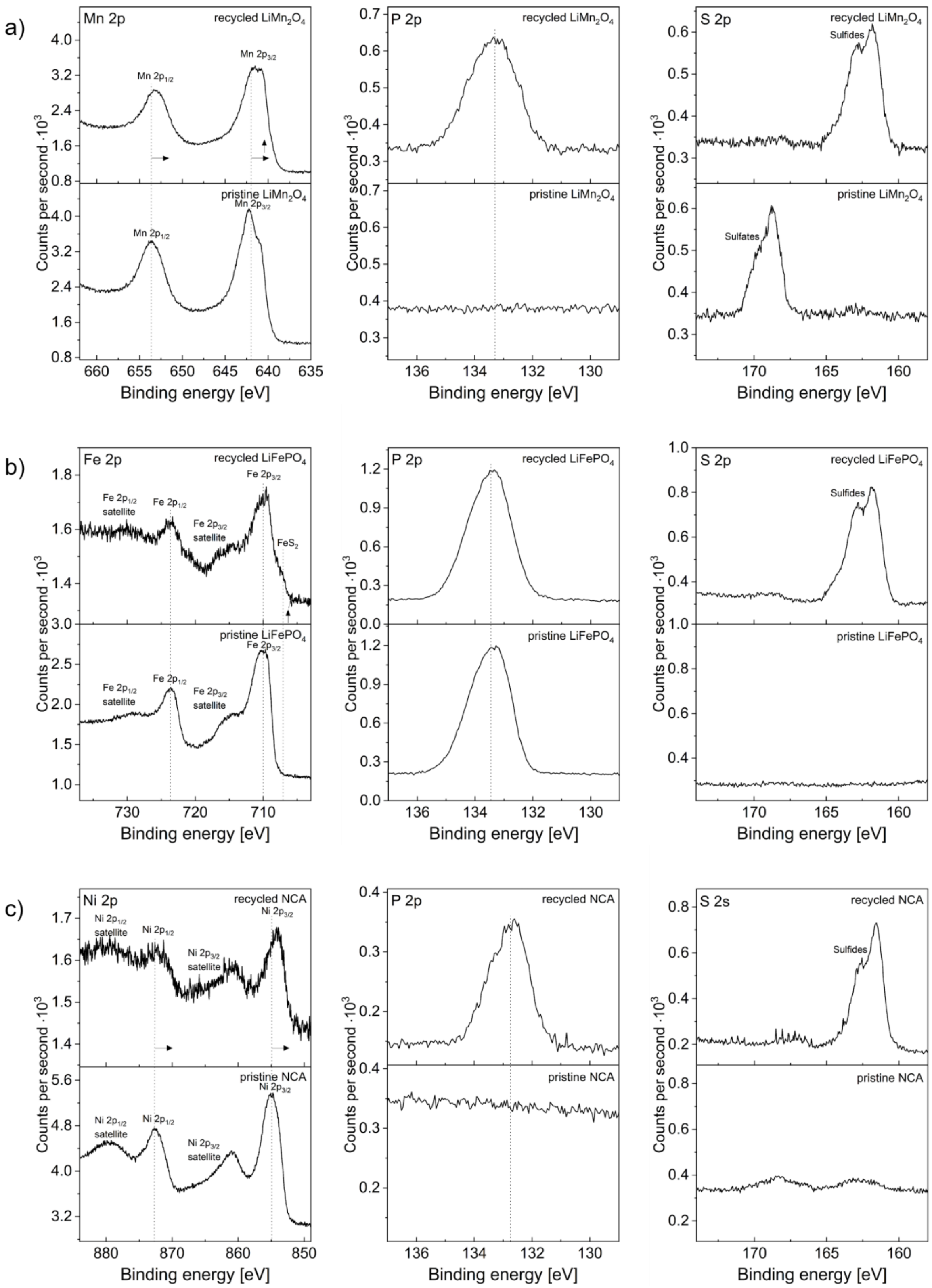

*Figure 5: Comparison of a) Mn 2p, P 2p and S 2s XP spectra of pristine and recycled LMO, b) Fe 2p, P 2p and S 2s XP spectra of pristine and recycled LFP, c) Ni 2p, P 2p and S 2s XP spectra of pristine and recycled NCA.*

Finally, the electrochemical behavior of the recycled electrode materials was examined in comparison to the pristine materials. For this, Li-ion batteries with an organic liquid electrolyte and tape-casted electrodes were assembled and cycled against metallic Li. Currently, the preparation of functional all-solid-state Li-ion batteries remains challenging and many factors can influence the electrochemical performance of such cells complicating the interpretation of results. Thus, to allow for a distinct identification of differences between the pristine and recycled materials tracing back to the dissolution-based separation process, a conventional cell set-up was chosen and CV measurements were performed. GCPL measurements were additionally conducted at different C-rates (Figure S8) and the corresponding rate capability was investigated (Figure S9). In general, the presence of phosphor and/or sulfur-containing species on the particles can be expected to have an influence on the cycling of the recycled materials as also observed in the performed electrochemical measurements.

For LCO and LTO, a good cycling behavior is found for both the pristine and recycled electrode material (Figure 6). The cycling performances are comparable to similar electrode materials reported in literature.[55, 59, 83, 84] This observation aligns with the found minimal morphological and compositional changes in both the electrode as well as the electrolyte fractions.

In contrast to this, pristine and recycled LMO, NMC811, LFP and NCA show significantly altered charging and discharging curves. In particular, for LFP and NCA, a deterioration of the cycling behavior is observed after the separation process. For LMO and NMC811, the main differences are found between the first cycle of the pristine and recycled electrode materials. Due to the presence of $Li_{1+x}Mn_2O_4$ in the recycled LMO, an additional oxidative peaks at ~ 3.3 V is related to the delithiation to $LiMn_2O_4$. The discharging potential limitation to 3.5 V does not allow for formation of fully discharged $Li_2Mn_2O_4$ during the subsequent discharging, instead only semi-discharged $LiMn_2O_4$ is formed and cycled in the following cycles. The comparison between the pristine and recycled NMC811 reveals a difference between the oxidation peak during the initial charging. While for pristine NMC811 a comparatively sharp peak is found at ~ 3.9 V, a broader peak shifted to ~ 4.0 V is observed for the recycled NMC811. Nevertheless, starting from the second cycle comparable charging and discharging curves are found for pristine and recycled LMO and NMC811, respectively.

In contrast du this, recycled LFP and NCA show significantly degradation. Though the pristine LFP shows already fairly low redox currents (note that the material was not coated with carbon which is in general required due to the poor electronic conductivity pf LFP), even worse behavior is found after the separation. This might be related to the formation of $FeS_2$ on the particle surfaces as shown by XPS. $FeS_2$ is known to have a low Li-ion conductivity.[85] For the recycled NCA, a strong decomposition under the formation of large fractions of NiO is observed. Though a thin layer of NiO has been reported to allow for a stabilization of the surface of Ni-rich cathodes, increased formation considerably increases the kinetic barrier for Li-ion diffusion.[42]

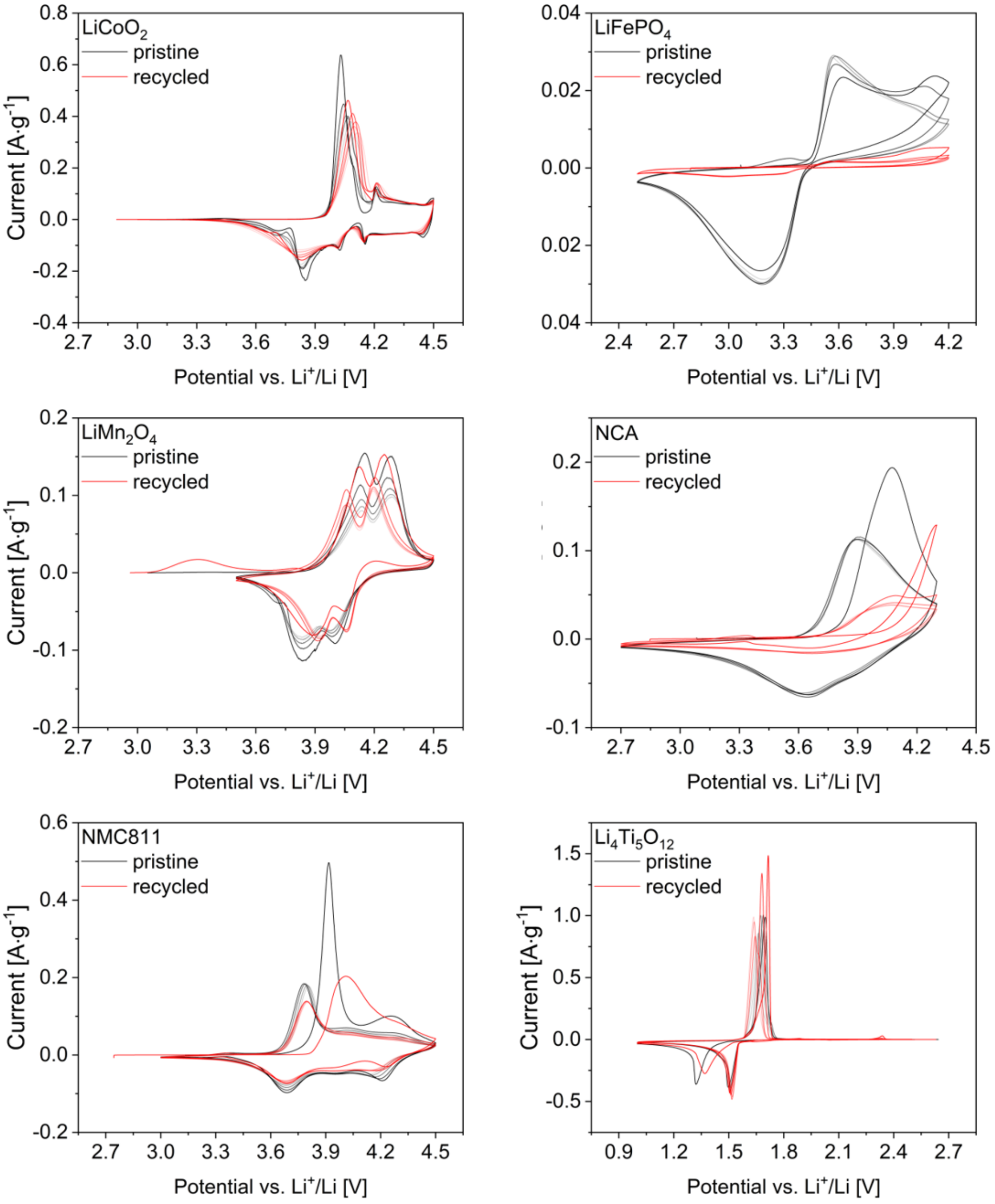

*Figure 6: Comparison of CV measurements of pristine and recycled electrode materials.*

# 4 Conclusions

In this article, it was shown that β-$Li_3PS_4$ can be separated from transition metal oxide electrode materials by dissolution in NMF. Though the separation results in a recovery of recycled β-$Li_3PS_4$ and electrode materials with relatively high purity, interactions between the materials during the dissolution process can cause strong changes in functional properties of both components. Interestingly, this behavior is hard to predict and different degradation reactions are observed. Though it can be assumed that the low stability of β-$Li_3PS_4$ could lead to redox interactions with electrode materials with late transition metals in high oxidation states, the impact on the electrolyte conductivity varies widely in dependence of the electrode material used. In addition, the electrochemical properties of the electrode materials can also change significantly. Due to the observed complexity of the ongoing interactions, the development of highly efficient recycling strategies enabling a closed-loop economy will likely require individual solutions depending on the cell chemistry since circular processes might be intrinsically different for different material combinations.

# 5 Supporting Information

The Supporting Information is available free of charge at https://xyz.org/....

Photographs of $Li_3PS_4$ containing NMF solutions after separation from different electrode materials; photograph of pristine and recycled $LiFePO_4$; table summarizing quantitative analysis, refined lattice parameters and amorphous phase fractions of pristine, recrystallized and recycled β-$Li_3PS_4$ after separation from different electrode materials; table summarizing quantitative analysis and refined lattice parameters of pristine and recycled electrode materials; table summarizing quantitative ICP-MS analysis of pristine and recycled β-$Li_3PS_4$ after separation from different electrode materials; Arrhenius plots Nyquist plots of pristine, recrystallized and recycled β-$Li_3PS_4$ after separation from different electrode materials measured at 25 °C and extended Nyquist plot of $LiNi_{0.85}Co_{0.1}Al_{0.05}O_2$; Bode plots of pristine, recrystallized and recycled β-$Li_3PS_4$ after separation from different electrode materials; chronoamperometry measurements of pristine, recrystallized and recycled β-$Li_3PS_4$ after separation from different electrode materials; table summarizing the EDX and ICP-MS analysis and corresponding chemical compositions of pristine and recycled electrode materials; comparison of additional XP spectra of pristine and recycled LMO, LFP and NCA; comparison of XP spectra of pristine and recycled NMC811; comparison between GCPL measurements pristine and recycled electrode materials; comparison between cycling performance pristine and recycled electrode materials.

# 6 Acknowledgements

This work has been funded by German federal state of Hessen (Hessen Agentur, HA-Project Number 848/20-08). ICP-MS instrumentation for this work was provided by the Elemental analysis group, with financial support from Saarland University and German Science Foundation (project number INST 256/553-1).

# 7 Authors' Contribution

KW, WE and OC conceived and designed the study. KW prepared the samples and performed measurements and analysis. KW wrote the manuscript. A.H. performed ICP-MS measurements on the electrolyte materials under the guidance of R. K. K.K. performed XPS measurements on the electrode materials under the guidance of U.S. All authors discussed and revised the work.

# 8 Conflicts of Interest

There are no conflicts of interest to declare.

Supporting Information

# Recycling of β-$Li_3PS_4$-based all-solid-state Li-ion batteries: Interactions of electrode materials and electrolyte in a dissolution process

Kerstin Wissel[a, b,*], Aaron Haben[c], Kathrin Küster[d], Ulrich Starke[d], Ralf Kautenburger[c], Wolfgang Ensinger[a] and Oliver Clemens[b]

[a] Technical University of Darmstadt, Institute for Materials Science, Materials Analysis, Alarich-Weiss-Straβe 2, 64287 Darmstadt, Germany

[b] University of Stuttgart, Institute for Materials Science, Chemical Materials Synthesis, Heisenbergstraβe 3, 70569 Stuttgart, Germany

[c] Saarland University, Inorganic Solid State Chemistry, Elemental Analysis, Campus C4 1, 66123 Saarbrücken, Germany

[d] Max Planck Institute for Solid State Research, Heisenbergstr. 1, 70569 Stuttgart, Germany

Corresponding Author:

Dr. Kerstin Wissel

Email: kerstin.wissel@imw.uni-stuttgart.de

Fax: +49 711 685 61963

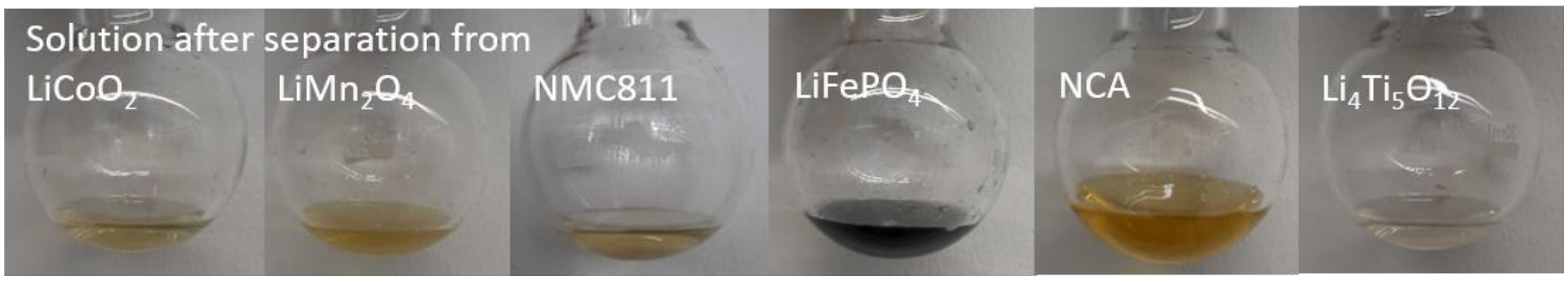

*Figure S 1: Photographs of $Li_3PS_4$ containing NMF solutions after separation from different electrode materials.*

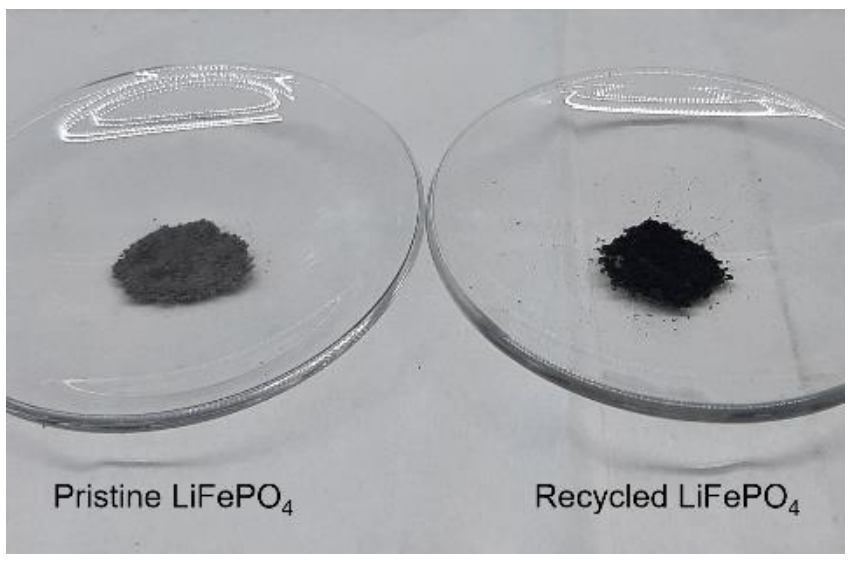

*Figure S 2: Photograph of pristine and recycled $LiFePO_4$.*

*Table S 1: Quantitative analysis, refined lattice parameters and amorphous phase fractions of pristine, recrystallized and recycled β-$Li_3PS_4$ after separation from different electrode materials. Phase fraction for which no numerical error is provided are subjected to an error of 1 to 2 wt.%.*

| | Phase | Phase fraction [wt.%] | Space group | a [Å] | b [Å] | c [Å] | Amorphous phase fraction [wt.%] |
|---|---|---|---|---|---|---|---|
| **pristine β-$Li_3PS_4$** | β-$Li_3PS_4$ | 96.75(6) | *Pnma* | 12.9959(4) | 8.0504(3) | 6.1430(2) | 7.9(7) |
| | $Li_2S$ | 3.25(6) | *Fm*-3*m* | 5.7158(3) | | | |
| **recrystallized β-$Li_3PS_4$** | β-$Li_3PS_4$ | 99.20(6) | *Pnma* | 12.9564(5) | 8.0961(3) | 6.1428(2) | 34.8(6) |
| | $Li_2S$ | 0.80(6) | *Fm*-3*m* | 5.724(1) | | | |
| **β-$Li_3PS_4$ separated from** | | | | | | | |
| **$LiCoO_2$** | β-$Li_3PS_4$ | 95 | *Pnma* | 12.9272(6) | 8.1049(4) | 6.1417(3) | 44 |
| | Unidentified phase(s) | 5 | | | | | |
| **$LiMn_2O_4$** | β-$Li_3PS_4$ | 100 | *Pnma* | 12.9137(11) | 8.1005(6) | 6.1361(4) | 52.5(6) |
| **NMC811** | β-$Li_3PS_4$ | 93 | *Pnma* | 12.9220(9) | 8.1054(5) | 6.1405(4) | 40 |
| | Unidentified phase(s) | 7 | | | | | |
| **$LiFePO_4$** | β-$Li_3PS_4$ | 93 | *Pnma* | 12.9006(9) | 8.1060(5) | 6.1320(4) | 47 |
| | Unidentified phase(s) | 7 | | | | | |
| **NCA** | β-$Li_3PS_4$ | 88 | *Pnma* | 12.9262(7) | 8.0961(4) | 6.1348(3) | 56 |
| | Unidentified phase(s) | 12 | | | | | |
| **$Li_4Ti_5O_{12}$** | β-$Li_3PS_4$ | 95 | *Pnma* | 12.9191(7) | 8.1016(4) | 6.1379(3) | 37 |
| | Unidentified phase(s) | 5 | | | | | |

*Table S 2: Quantitative analysis and refined lattice parameters of pristine and recycled electrode materials.*

| | Phase | Phase fraction [%] | Space group | a [Å] | b [Å] | c [Å] | α [°] | β [°] | γ [°] |
|---|---|---|---|---|---|---|---|---|---|
| **pristine $LiCoO_2$** | $LiCoO_2$ | 100 | *R*-3*m* | 2.813772(12) | | 14.05267(15) | | | |
| **recycled $LiCoO_2$** | $LiCoO_2$ | 100 | *R*-3*m* | 2.813846(15) | | 14.05457(18) | | | |
| **pristine $LiMn_2O_4$** | $LiMn_2O_4$ | 100 | *Fd*-3*m* | 8.239496(19) | | | | | |
| **recycled $LiMn_2O_4$** | $LiMn_2O_4$ | 83.13(13) | *Fd*-3*m* | 8.2167(8) | | | | | |
| | $Li_2Mn_2O_4$ | 16.87(13) | $I4_1/amd$ | 5.6369(5) | | 9.2163(9) | | | |
| **pristine NMC811** | NMC811 | 100 | *R*-3*m* | 2.877537(17) | | 14.20632(18) | | | |
| **recycled NMC811** | NMC811 | 100 | *R*-3*m* | 2.87825(3) | | 14.2081(4) | | | |
| **pristine $LiFePO_4$** | $LiFePO_4$ | 87.1(2) | Pnma | 10.32995(10) | 6.00742(6) | 4.69210(5) | | | |
| | $Fe_2P_2O_7$ | 12.9(2) | C-1 | 6.6426(10) | 8.4632(15) | 4.4868(8) | 89.997(10) | 103.796(12) | 92.697(17) |
| **recycled $LiFePO_4$** | $LiFePO_4$ | 88.6(3) | *Pnma* | 10.3371(2) | 6.01098(14) | 4.69491(11) | | | |
| | $Fe_2P2O_7$ | 11.4(3) | *C*-1 | 6.6453(14) | 8.465(2) | 4.4891(12) | 90.036(15) | 103.767(18) | 92.62(2) |
| **pristine NCA** | NCA | 100 | *R*-3*m* | 2.86858(7) | | 14.1759(4) | | | |
| **recycled NCA** | NCA | 76.8(3) | *R*-3*m* | 2.86616(3) | | 14.1921(8) | | | |
| | NiO | 23.2(3) | *Fm*-3*m* | 2.9748(6) | | 14.1921(8) | | | |
| **pristine $Li_4Ti_5O_{12}$** | $Li_4Ti_5O_{12}$ | 98.38(5) | *Fd*-3*m* | 8.35582(3) | | | | | |
| | $TiO_2$ | 1.62(5) | $P4_2/mnm$ | 4.5823(4) | | 2.9547(5) | | | |
| **recycled $Li_4Ti_5O_{12}$** | $Li_4Ti_5O_{12}$ | 98.67(6) | *Fd*-3*m* | 8.35493(5) | | | | | |
| | $TiO_2$ | 1.33(6) | $P4_2/mnm$ | 4.5814(5) | | 2.9561(7) | | | |

*Table S 3: Quantitative ICP-MS analysis of pristine and recycled β-$Li_3PS_4$ after separation from different electrode materials. S contents were not determined.*

| | Mass [mg] | Volume [ml] | Molar concentration [µmol/l] | | | | | | | | Nominal molar concentration [µmol/l] | |
|---|---|---|---|---|---|---|---|---|---|---|---|---|
| | | | Li | Al | P | Ti | Mn | Fe | Co | Ni | Li | P |
| **pristine β-$Li_3PS_4$** | 4.1 | 10 | 5788.6 | ND | 1872.2 | 0.096 | 0.033 | ND | 0.039 | 0.051 | 6832.2 | 2277.4 |
| **β-$Li_3PS_4$ separated from** | | | | | | | | | | | | |
| **$LiCoO_2$** | 4.0 | 10 | 7646.9 | ND | 2433.4 | 0.070 | 0.019 | ND | 0.150 | ND | 6665.6 | 2221.9 |
| **$LiMn_2O_4$** | 4.0 | 10 | 6166.9 | ND | 2034.4 | 0.066 | 2.741 | ND | 0.004 | 0.017 | 7332.1 | 2444.0 |
| **NMC811** | 4.4 | 10 | 6736.7 | ND | 2090.4 | 0.041 | 0.082 | ND | 0.024 | 0.144 | 6736.7 | <1.853 |
| **$LiFePO_4$** | 4.0 | 10 | 7450.8 | ND | 2455.5 | 0.066 | 0.034 | 1.623 | 0.014 | 0.039 | 6665.6 | 2221.9 |
| **NCA** | 4.0 | 10 | 5679.6 | ND | 1838.4 | 0.030 | 0.049 | ND | 0.243 | 0.946 | 6665.6 | 2221.9 |
| **$Li_4Ti_5O_{12}$** | 4.0 | 10 | 6626.0 | ND | 2134.4 | 0.075 | 0.025 | ND | 0.005 | ND | 6665.6 | 2221.9 |

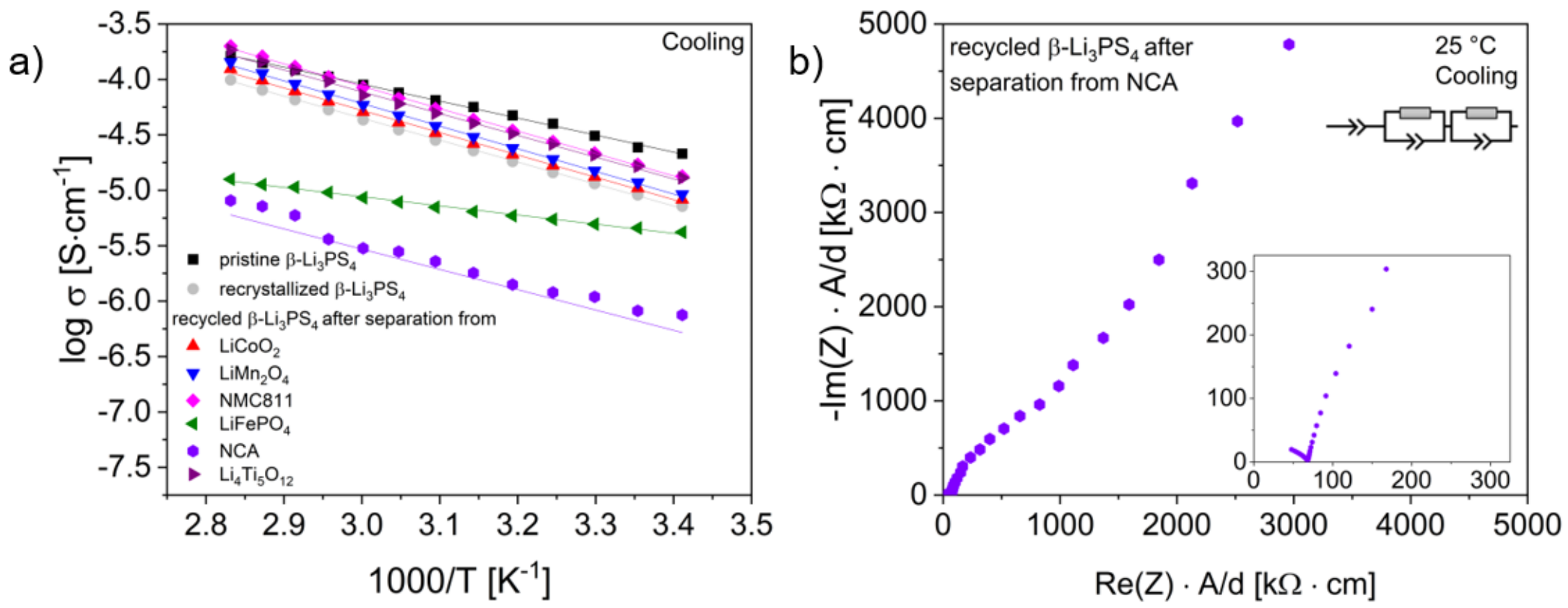

*Figure S 3: a) Arrhenius plots of pristine, recrystallized and recycled β-$Li_3PS_4$ after separation from different electrode materials determined during the cooling sweep of the EIS measurement. b) Extended Nyquist plot of recycled β-$Li_3PS_4$ after separation from NCA.*

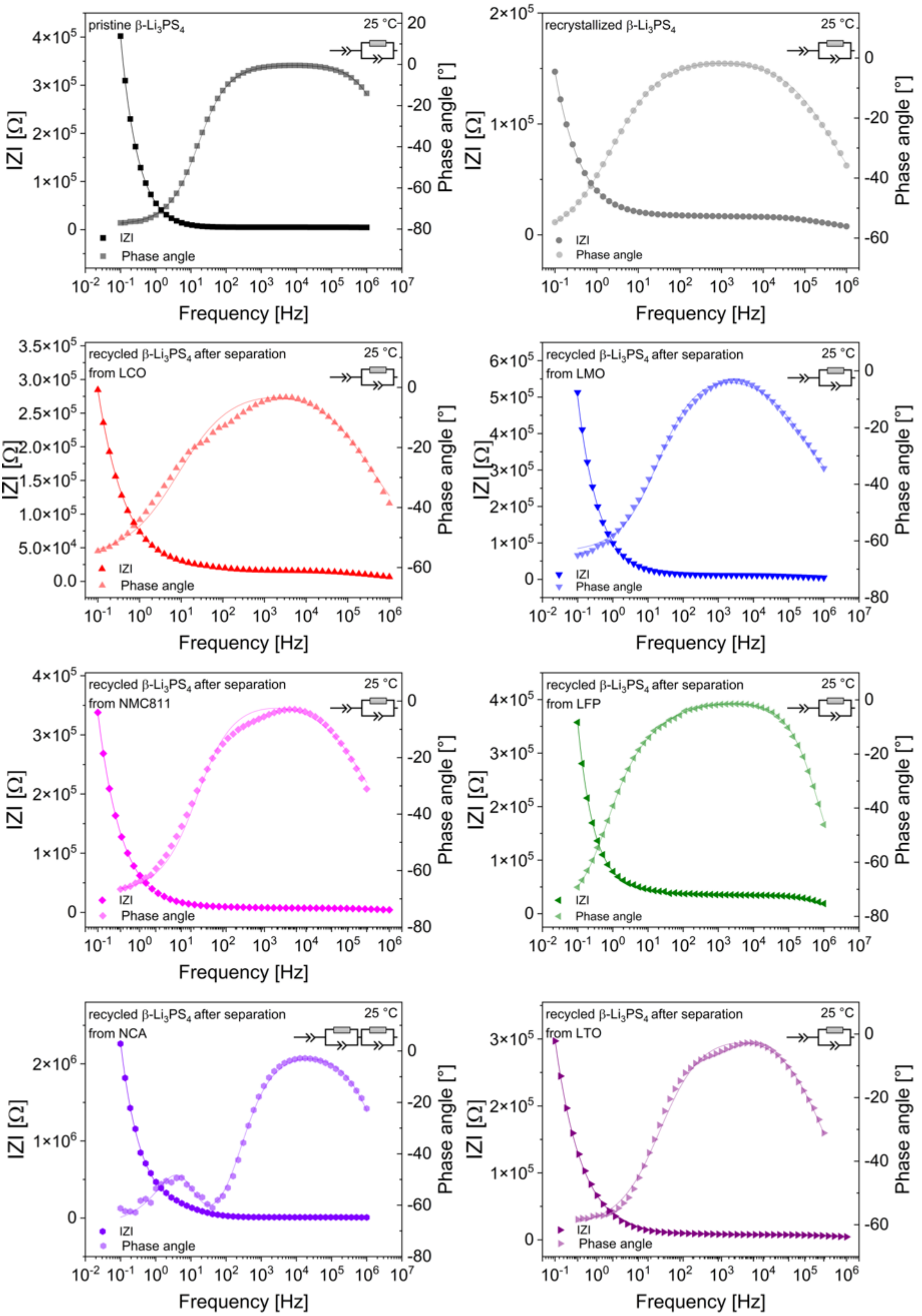

*Figure S 4: Bode plots of pristine, recrystallized and recycled β-$Li_3PS_4$ after separation from different electrode materials measured at 25 °C with corresponding fits. In addition, the equivalent circuits used for fitting are provided.*

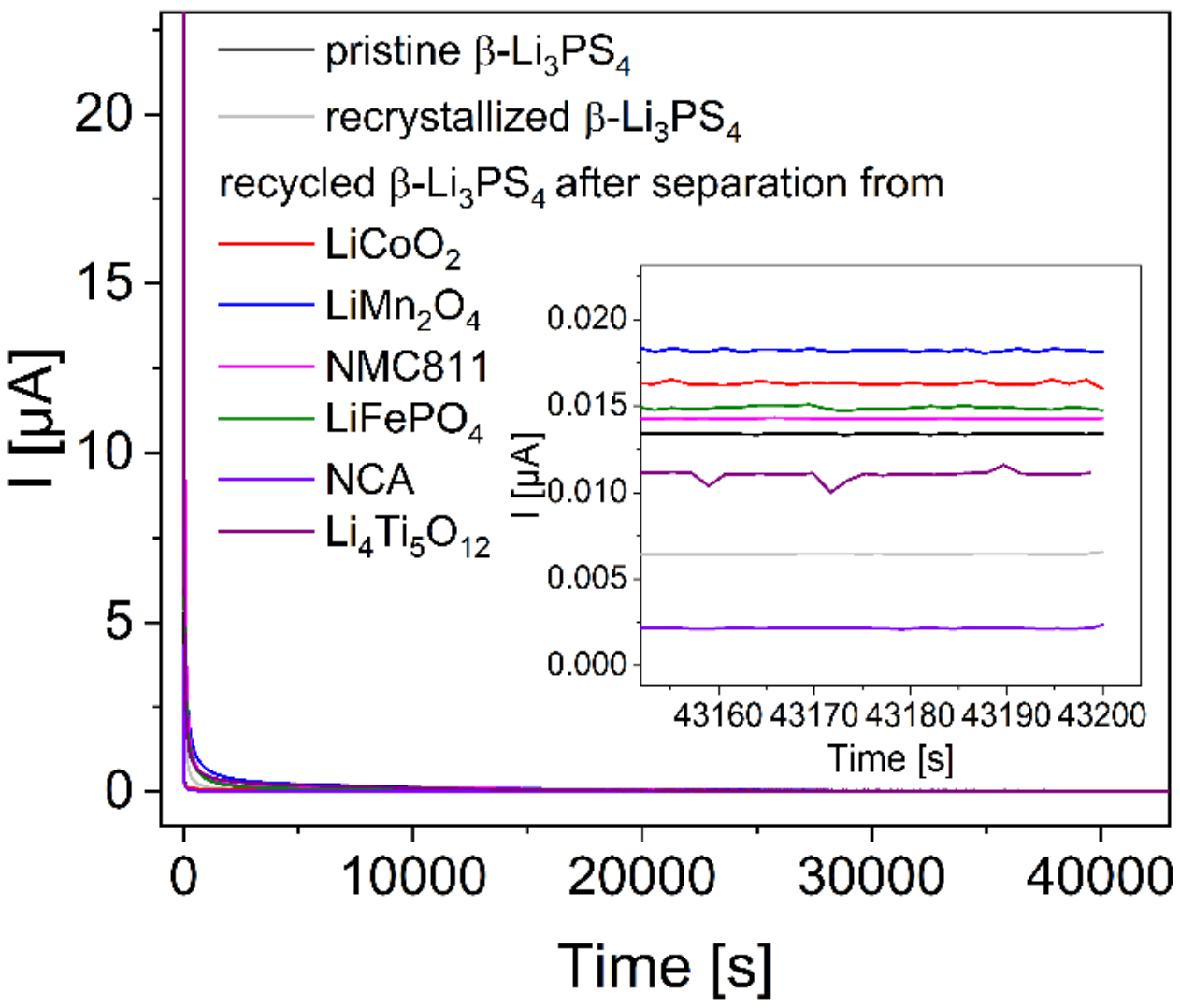

*Figure S 5: Chronoamperometry measurements of pristine, recrystallized and recycled β-$Li_3PS_4$ after separation from different electrode materials.*

*Table S 4: EDX analysis and corresponding chemical compositions of pristine and recycled electrode materials. Each sample was measured at three different spots. For comparison, the compositions based on ICP-MS measurements are given. *: S contents were not determined.*

| | At.% of Co [%] | At.% of Mn [%] | At.% of Ni [%] | At.% of Fe [%] | At.% of Al [%] | At.% of Ti [%] | At.% of O [%] | At.% of P [%] | At.% of S [%] | Composition based on EDX | Composition based on ICP-MS* |
|---|---|---|---|---|---|---|---|---|---|---|---|
| **pristine $LiCoO_2$** | 26.64 | | | | | | 63.36 | | | $Li_xCoO_{1.73}$ | $Li_{1.06}CoO_x$ |
| | 33.00 | | | | | | 67.00 | | | $Li_xCoO_{2.03}$ | |
| | 31.33 | | | | | | 68.76 | | | $Li_xCoO_{2.19}$ | |
| **recycled $LiCoO_2$** | 35.24 | | | | | | 63.96 | 0.31 | 0.48 | $Li_xCoO_{1.81}$ $P_{0.009}S_{0.013}$ | $Li_{1.02}CoO_x$ $P_{0.001}$ |
| | 37.63 | | | | | | 61.40 | 0.42 | 0.54 | $Li_xCoO_{1.63}$ $P_{0.011}S_{0.014}$ | |
| | 40.57 | | | | | | 58.83 | 0.22 | 0.39 | $Li_xCoO_{1.45}$ $P_{0.005}S_{0.010}$ | |
| **pristine $LiMn_2O_4$** | | 24.89 | | | | | 75.11 | | | $Li_xMn_2O_{6.04}$ | $Li_{0.75}Mn_2O_x$ |
| | | 44.79 | | | | | 55.21 | | | $Li_xMn_2O_{2.47}$ | |
| | | 37.72 | | | | | 62.28 | | | $Li_xMn_2O_{3.30}$ | |
| **recycled $LiMn_2O_4$** | | 34.12 | | | | | 65.14 | 0.34 | 0.35 | $Li_xMn_2O_{3.81}$ $P_{0.02}S_{0.02}$ | $Li_{0.85}Mn_2O_x$ $P_{0.001}$ |
| | | 31.20 | | | | | 67.94 | 0.43 | 0.43 | $Li_xMn_2O_{4.81}$ $P_{0.03}S_{0.03}$ | |
| | | 29.14 | | | | | 70.08 | 0.33 | 0.45 | $Li_xMn_2O_{4.81}$ $P_{0.02}S_{0.03}$ | |
| **pristine NMC811** | 4.09 | 3.26 | 30.29 | | | | 62.37 | | | $Li_xNi_{0.8}Mn_{0.10}Co_{0.11}O_{1.65}$ | $Li_{0.93}Ni_{0.8}Mn_{0.11}$ $Co_{0.10}O_x$ |
| | 3.60 | 3.14 | 30.74 | | | | 62.51 | | | $Li_xNi_{0.8}Mn_{0.08}Co_{0.09}O_{1.63}$ | |
| | 4.05 | 2.87 | 29.47 | | | | 63.62 | | | $Li_xNi0.8Mn_{0.08}Co_{0.11}O_{1.73}$ | |
| **recycled NMC811** | 3.92 | 3.38 | 28.48 | | | | 63.57 | 0.24 | 0.42 | $Li_xNi_{0.8}Mn_{0.10}Co_{0.11}O_{1.79}$ $P_{0.007}S_{0.012}$ | $Li_{0.61}Ni_{0.8}Mn_{0.10}$ $Co_{0.09}O_x$ $P_{0.003}$ |
| | 4.11 | 3.04 | 30.96 | | | | 61.29 | 0.30 | 0.30 | $Li_xNi_{0.8}Mn_{0.08}Co_{0.11}O_{1.58}$ $P_{0.008}S_{0.008}$ | |
| | 3.29 | 2.18 | 23.77 | | | | 65.28 | 3.29 | 2.18 | $Li_xNi0.8Mn0.07Co_{0.11}O_{2.20}$ $P_{0.111}S_{0.073}$ | |
| **pristine $LiFePO_4$** | | | | 19.10 | | | 66.77 | 14.13 | | $Li_xFe_1P_{0.74}O_{3.50}$ | $Li_{0.98}Fe_1P_{1.45}O_x$ |
| | | | | 17.80 | | | 69.35 | 12.85 | | $Li_xFe_1P_{0.72}O_{3.90}$ | |
| | | | | 16.63 | | | 69.63 | 13.75 | | $Li_xFe_1P_{0.83}O_{4.19}$ | |
| **recycled $LiFePO_4$** | | | | 17.73 | | | 69.67 | 11.79 | 0.82 | $Li_xFe_1P_{0.67}O_{3.93}$ $S_{0.046}$ | $Li_{1.27}Fe_1P_{1.07}O_x$ |
| | | | | 19.43 | | | 66.87 | 12.24 | 1.45 | $Li_xFe_1P_{0.63}O_{3.44}$ $S_{0.075}$ | |
| | | | | 20.97 | | | 63.10 | 14.99 | 0.94 | $Li_xFe_1P_{0.72}O_{3.01}$ $S_{0.045}$ | |
| **pristine NCA** | 6.76 | | 34.27 | | 1.86 | | 57.12 | | | $Li_xNi_{0.8}Co_{0.16}Al_{0.04}O_{1.33}$ | $Li_{1.02}Ni_{0.8}Co_{0.13}$ $Al_{0.05}O_x$ |
| | 4.63 | | 25.58 | | 1.32 | | 68.47 | | | $Li_xNi_{0.8}Co_{0.14}Al_{0.04}O_{2.14}$ | |
| | 6.03 | | 30.12 | | 1.47 | | 62.38 | | | $Li_xNi_{0.8}Co_{0.16}Al_{0.03}O_{1.66}$ | |
| **recycled NCA** | 4.60 | | 23.43 | | 2.32 | | 63.62 | 1.85 | 4.17 | $Li_xNi_{0.8}Co_{0.16}Al_{0.08}O_{2.17}$ $P_{0.063}S_{0.142}$ | $Li_{1.09}Ni_{0.8}Co_{0.14}$ $Al_{0.04}O_x$ $P_{0.023}$ |
| | 4.93 | | 25.46 | | 1.75 | | 64.94 | 0.74 | 2.18 | $Li_xNi_{0.8}Co_{0.16}Al_{0.04}O_{1.33}$ $P_{0.023}S_{0.069}$ | |
| | 4.37 | | 22.11 | | 1.62 | | 66.09 | 1.76 | 4.05 | $Li_xNi_{0.8}Co_{0.16}Al_{0.04}O_{1.33}$ $P_{0.064}S_{0.146}$ | |
| **pristine $Li_4Ti_5O_{12}$** | | | | | | 33.52 | 66.48 | | | $Li_xTi_5O_{9.91}$ | - |
| | | | | | | 27.75 | 72.25 | | | $Li_xTi_5O_{13.02}$ | |
| | | | | | | 23.62 | 76.38 | | | $Li_xTi_5O_{16.17}$ | |
| **recycled $Li_4Ti_5O_{12}$** | | | | | | 32.99 | 65.95 | 0.45 | 0.61 | $Li_xTi_5O_{10.00}$ $P_{0.068}S_{0.047}$ | - |
| | | | | | | 32.13 | 66.10 | 0.45 | 1.32 | $Li_xTi_5O_{10.29}$ $P_{0.070}S_{0.100}$ | |
| | | | | | | 29.66 | 68.95 | 0.32 | 1.07 | $Li_xTi_5O_{11.62}$ $P_{0.053}S_{0.078}$ | |

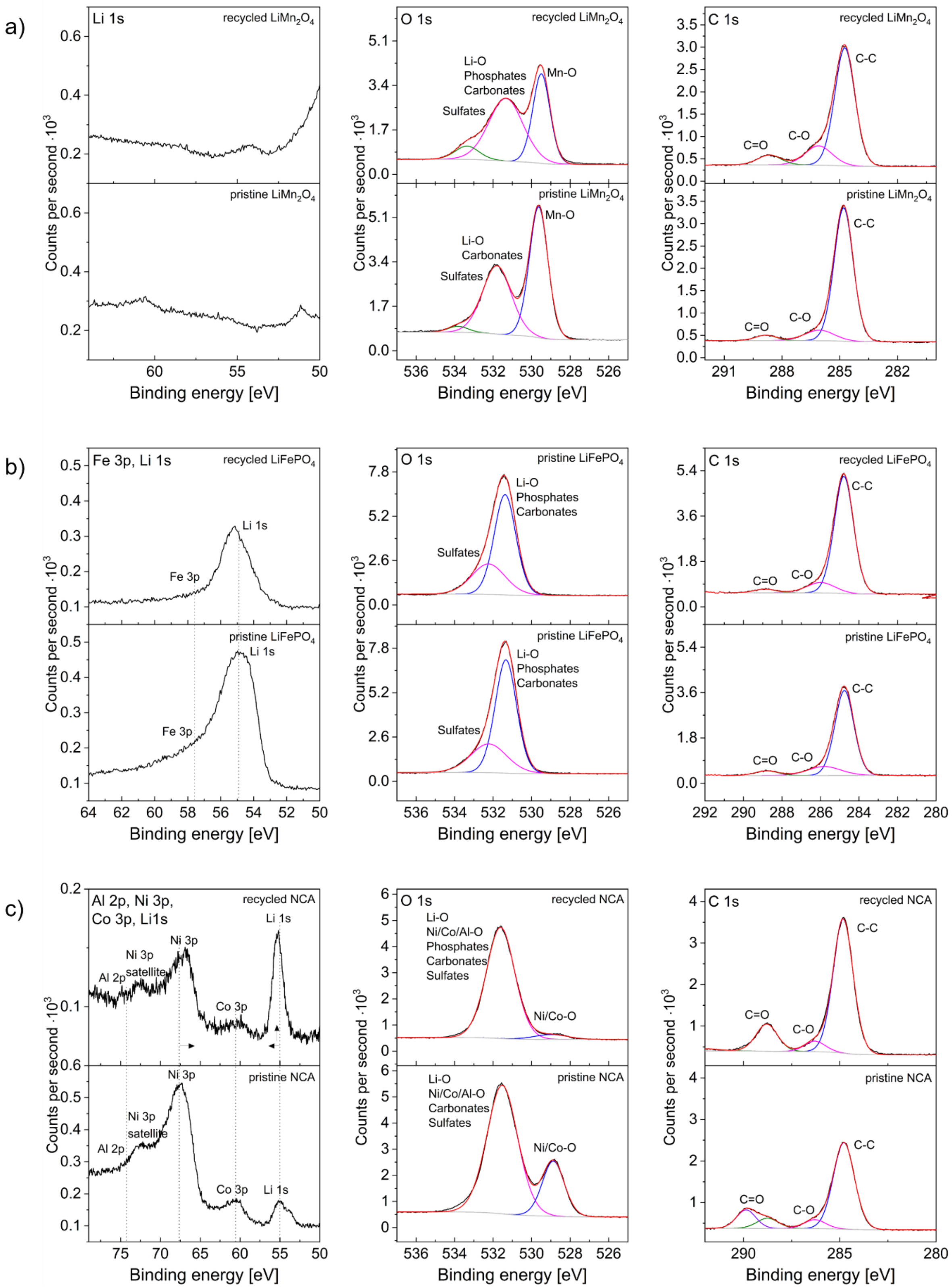

*Figure S 6: Comparison of a) Li 1s, O 1s and C 1s XP spectra of pristine and recycled LMO, b) Fe 3p and Li 1s, O 1s and C 1s XP spectra of pristine and recycled LFP, c) Al 2p, Ni 3p, Co 3p and Li 1s, O1s and C 1s XP spectra of pristine and recycled NCA.*

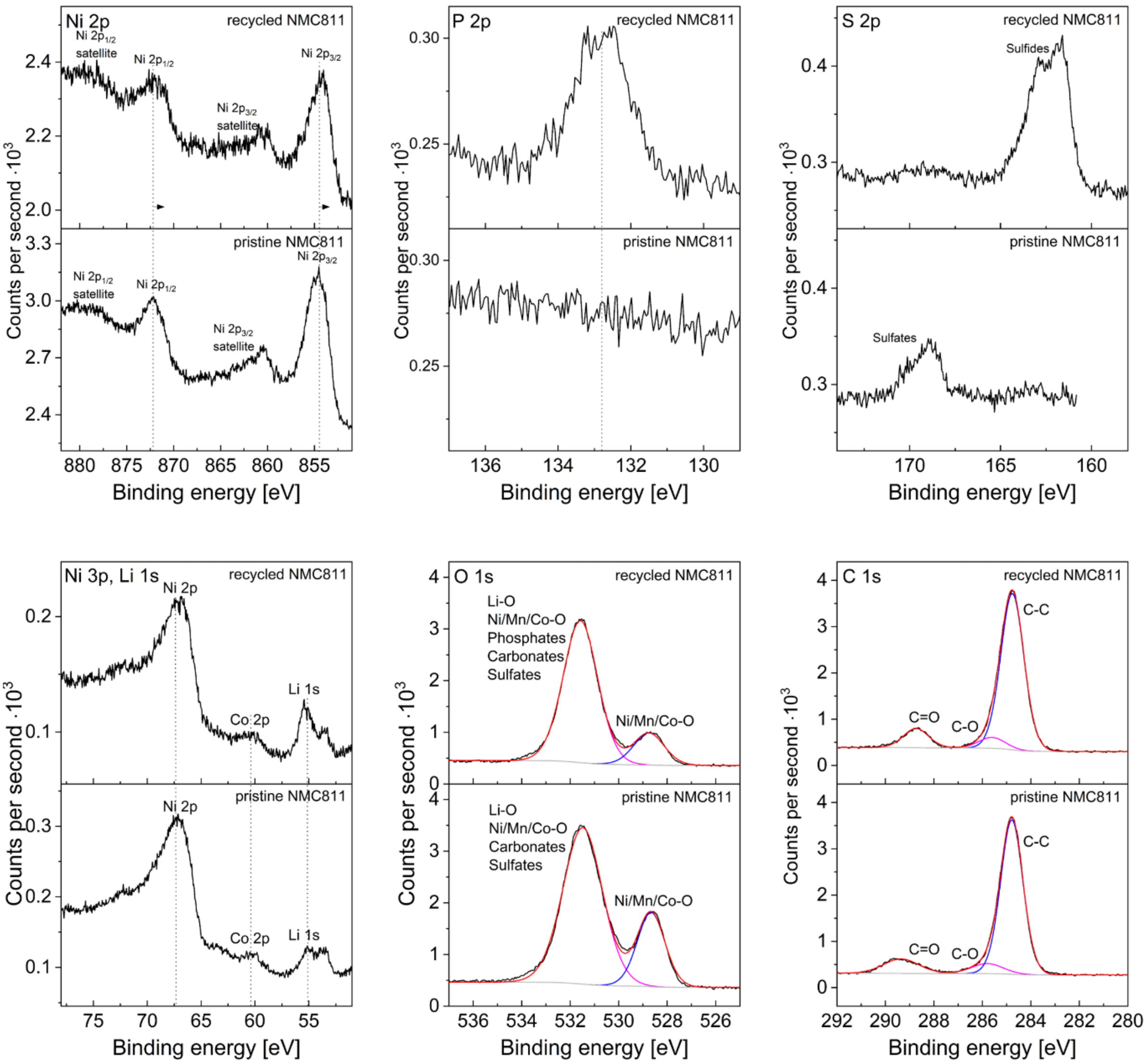

*Figure S 7: Comparison of Ni 2p, P 2p, S 2p, Ni 3p and Li 1s, O1s and C 1s XP spectra of pristine and recycled NMN811.*

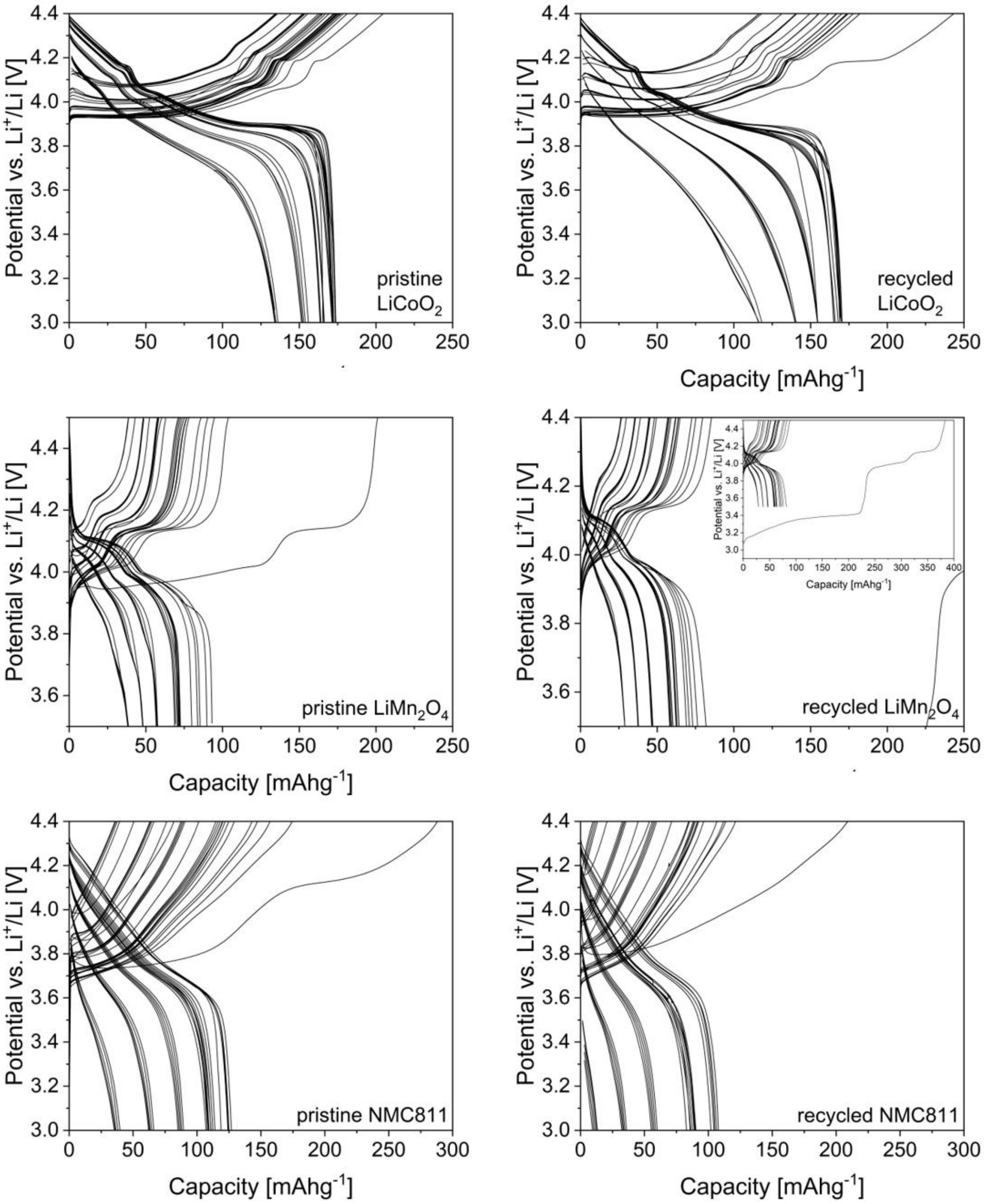

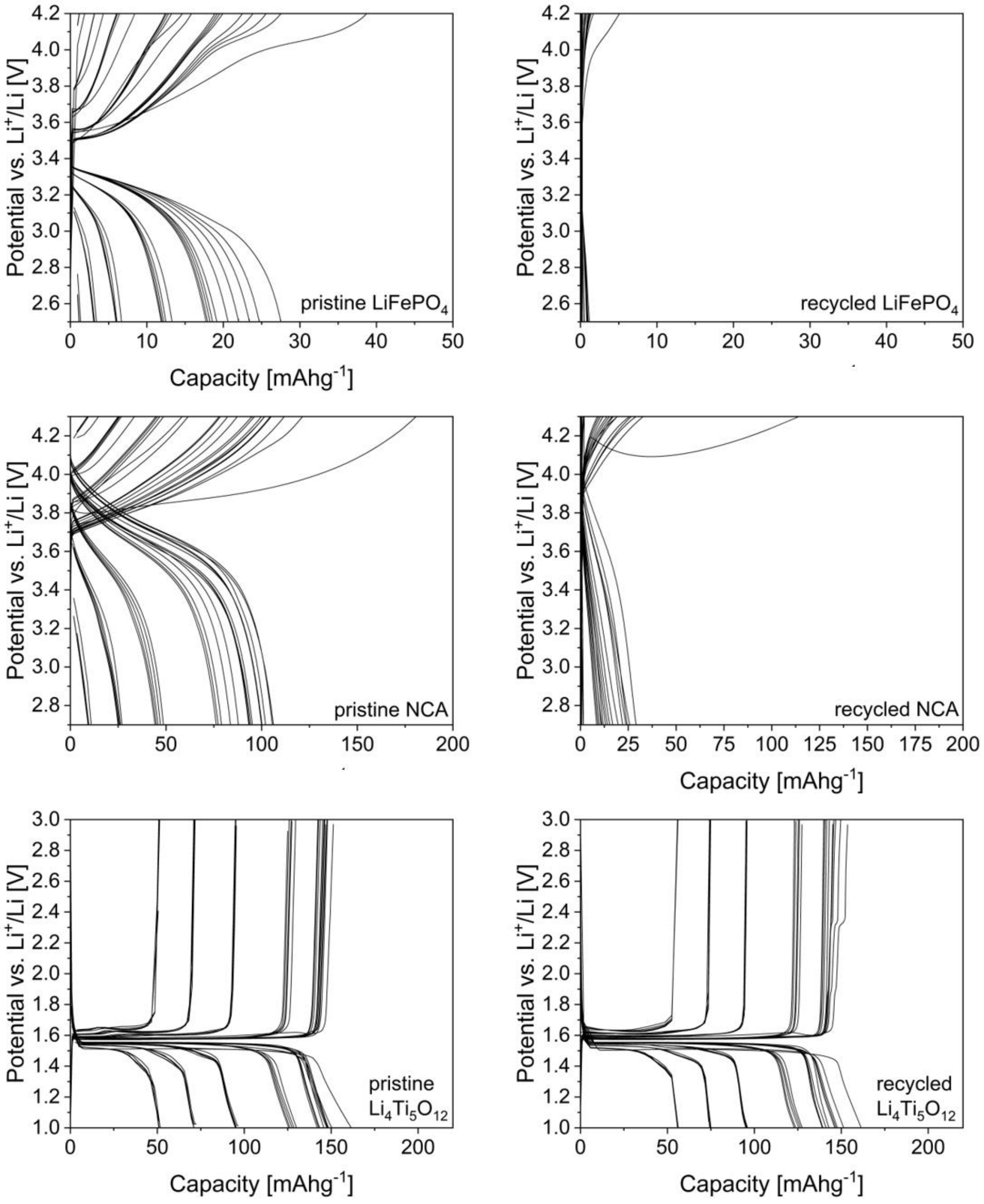

*Figure S 8: Comparison between GCPL measurements pristine and recycled electrode materials.*

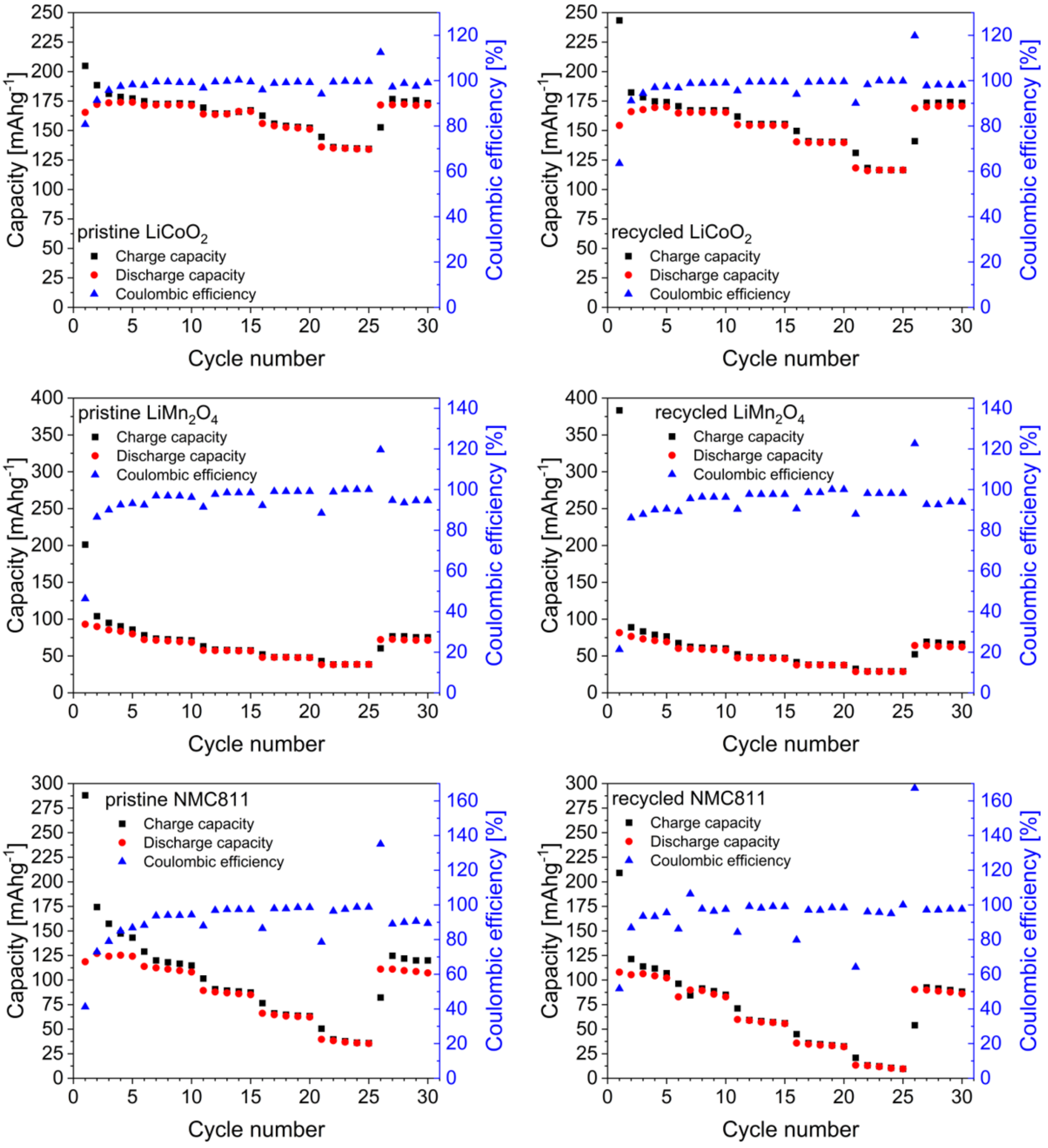

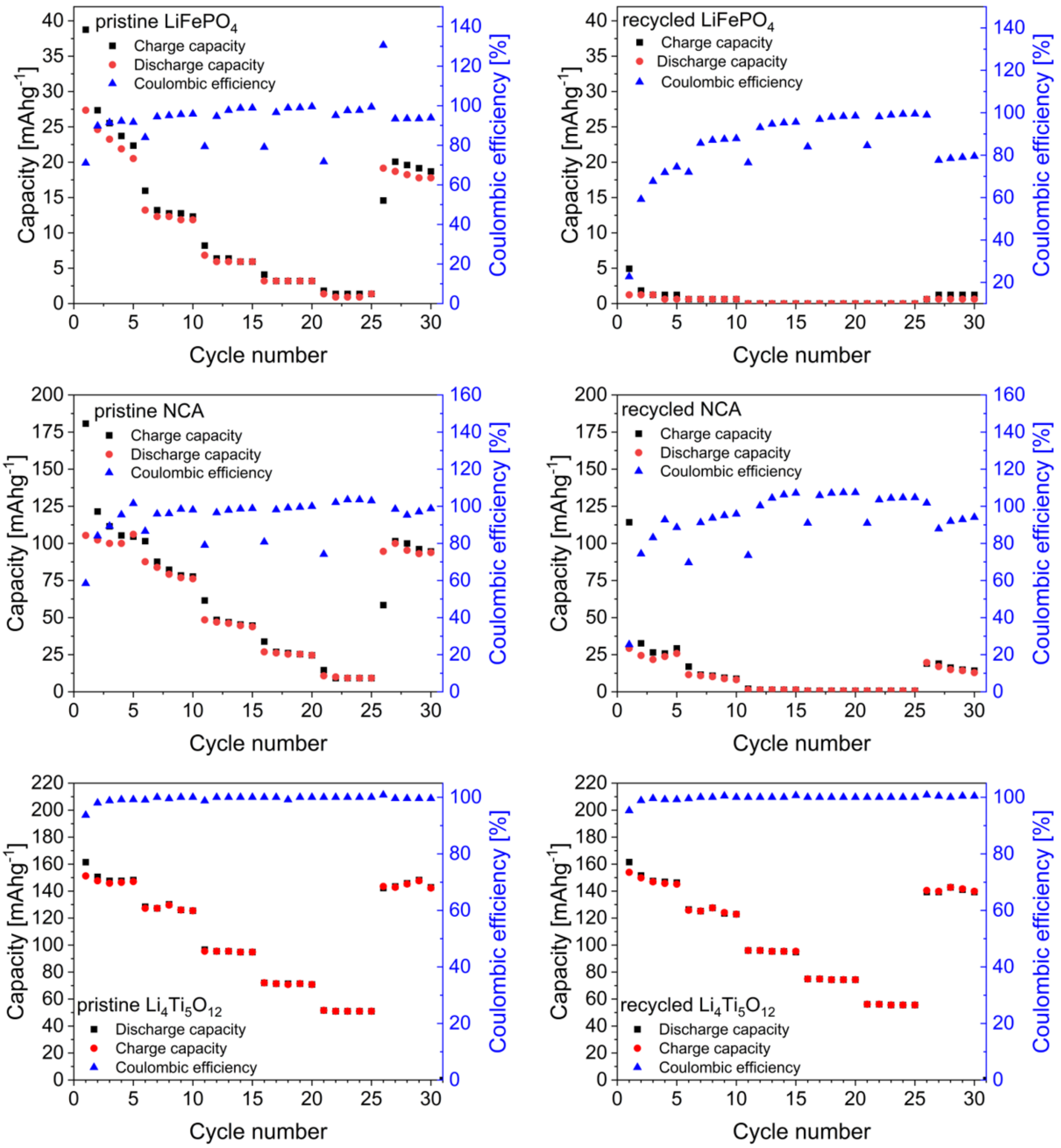

*Figure S 9: Comparison between cycling performance pristine and recycled electrode materials.*